\documentclass[amsmath,amssymb,aps,10pt,prd,letterpaper,nofootinbib,balancelastpage,notitlepage,superscriptaddress,twocolumn,floatfix,preprintnumbers]{revtex4-2}
\usepackage{graphicx}	
\usepackage{ragged2e}	
\usepackage{bm}		    
\usepackage[sort&compress]{natbib}	 	
	\setcitestyle{square,numbers,comma}	
\usepackage[colorlinks=true,urlcolor=blue,linkcolor=red,citecolor=red]{hyperref}	
\newcommand{\tabref}[2][]{Tab{#1}.~\ref{tab:#2}}		
\newcommand{\figref}[2][]{Fig{#1}.~\ref{fig:#2}}		
\newcommand{\secref}[2][]{Sec{#1}.~\ref{sec:#2}}		
\newcommand{\appref}[2][x]{Appendi{#1}~\ref{app:#2}}	
\renewcommand{\eqref}[2][]{Eq{#1}.~(\ref{eq:#2})}		
\newcommand{\eqrefRange}[2]{Eqs.~(\ref{eq:#1})--(\ref{eq:#2})}		
\newcommand{\citeR}[2][]{Ref{#1}.~\cite{#2}}			
\newcommand{\eV}{\,\textrm{eV}}
\newcommand{\DM}{\mathrm{DM}}
\newcommand{\Hz}{\,\textrm{Hz}}
\DeclareMathOperator{\IM}{\mathrm{Im}}
\newcommand{\LL}{\mathcal{L}}
\newcommand{\nl}{\nonumber \\ & \quad }					
\newcommand{\phihat}{\bm{\hat{\phi}}}
\DeclareMathOperator{\RE}{\mathrm{Re}}
\newcommand{\rhat}{\bm{\hat{r}}}
\newcommand{\thetahat}{\bm{\hat{\theta}}}

\begin{document}

\title{Search for ultralight dark matter in the SuperMAG high-fidelity dataset}
\date{\today}
\author{Matt Friel}
\affiliation{Applied Physics Laboratory, Johns Hopkins University, Laurel, MD 20723, USA}
\author{Jesper W.~Gjerloev}
\affiliation{Applied Physics Laboratory, Johns Hopkins University, Laurel, MD 20723, USA}
\author{Saarik Kalia}
\email{kalias@umn.edu}
\affiliation{School of Physics \& Astronomy, University of Minnesota, Minneapolis, MN 55455, USA}
\author{Alvaro Zamora}
\affiliation{Kavli Institute for Particle Astrophysics \& Cosmology, Stanford University, Stanford, CA 94305, USA}

\preprint{FERMILAB-PUB-24-0546-SQMS}

\begin{abstract}
Ultralight dark matter, such as kinetically mixed dark-photon dark matter (DPDM) or axionlike-particle dark matter (axion DM), can source an oscillating magnetic-field signal at Earth's surface.  Previous work searched for this signal in a publicly available dataset of global magnetometer measurements maintained by the SuperMAG collaboration.  This ``low-fidelity" dataset reported measurements with a 1-minute time resolution, allowing the search to set leading direct constraints on DPDM and axion DM with Compton frequencies $f_\DM\leq1/(1\,\mathrm{min})$ [corresponding to masses $m_\DM\leq7\times10^{-17}\eV$].  More recently, a dedicated experiment undertaken by the SNIPE Hunt collaboration has also searched for this same signal at higher frequencies $f_\DM\geq0.5\Hz$ (or $m_\DM\geq2\times10^{-15}\eV$).  In this work, we search for this signal of ultralight DM in the SuperMAG ``high-fidelity" dataset, which features a 1-second time resolution, allowing us to probe the gap in parameter space between the low-fidelity dataset and the SNIPE Hunt experiment.  The high-fidelity dataset exhibits lower geomagnetic noise than the low-fidelity dataset and features more data than the SNIPE Hunt experiment, making it a powerful probe of ultralight DM.  Our search finds no robust DPDM or axion DM candidates.  We set constraints on DPDM and axion DM parameter space for $10^{-3}\Hz\leq f_\DM\leq0.98\Hz$ (or $4\times10^{-18}\eV\leq m_\DM\leq4\times10^{-15}\eV$).  Our results are the leading direct constraints on both DPDM and axion DM in this mass range, and our DPDM constraint surpasses the leading astrophysical constraint in a narrow range around $m_{A'}\approx2\times10^{-15}\eV$.
\end{abstract}

\maketitle

\section{Introduction}
\label{sec:introduction}

Discerning the fundamental nature of dark matter (DM) remains one of the most prominent open problems in modern physics.  A number of potential DM candidates have been proposed, but one class that has recently received significant attention is ultralight bosonic DM~\cite{Arias:2012az,kimball2022search}.  This class includes DM candidates with masses $m_\DM\lesssim1\eV$,%
\footnote{We work in natural units where $\hbar=c=1$.}
which have correspondingly large number densities and so behave as classical fields~\cite{lin2018self,Centers:2019dyn,cheong2024}.  Two such candidates that have garnered particular interest are dark-photon dark matter (DPDM)~\cite{Holdom:1986ag,cvetivc1996implications,Nelson:2011sf,Graham:2015rva} and axionlike-particle dark matter (henceforth simply axion DM)~\cite{Preskill:1982cy,Abbott:1982af,Dine:1982ah,Svrcek:2006yi,Arvanitaki:2009fg}.  Both of these candidates may exhibit couplings with electromagnetism~\cite{Holdom:1986ag,Sikivie:1983ip},%
\footnote{In the context of grand unified theories (GUTs) or heterotic string theories, an axionlike particle that couples to photons should also possess a coupling to gluons, which would contribute to its mass~\cite{Agrawal_2022,Agrawal_2024}.  The discovery of a light axionlike particle coupled to photons would therefore rule out many simple GUT and string theory models.}
and a number of experimental~\cite{ehret2010new,Wagner:2010mi,Redondo:2010dp,Horns:2012jf,Betz_2013,Graham:2014sha,Chaudhuri:2014dla,graham2016dark,Caldwell:2016dcw,Anastassopoulos:2017ftl,Baryakhtar:2018doz,armengaud2019physics,Lawson:2019brd,gramolin2021search,Andrianavalomahefa:2020ucg,Salemi:2021gck,Chiles_2022,haystaccollaboration2023new,romanenko2023new}, astrophysical~\cite{Dubovsky_2015,Wadekar_2021,Sisk_Reynes_2021,Dolan_2022,Hoof_2023,li2023,dolan2023,an2024,manzari2024}, and cosmological probes~\cite{Cadamuro_2012,McDermott_2020,Caputo_2020,Ferguson_2022,Langhoff_2022,Adachi_2023} have been utilized to constrain their parameter spaces.

Few experiments searching for these ultralight DM candidates are sensitive to the very low mass parameter space $m_\DM\lesssim10^{-12}\eV$.  This is because the signal in these experiments typically scales with the size of the experiment $L$, when $\lambda_\DM\gg L$ (where $\lambda_\DM=2\pi/m_\DM$ is the Compton wavelength of the DM).  One notable example is a series of experimental searches that utilize Earth as a transducer for ultralight DM~\cite{Fedderke:2021rys,Fedderke:2021iqw,Arza_2022,Sulai_2023,Bloch_2024}.  In such contexts, the effective size of the experiment is the radius of Earth $R$, which leads to a large enhancement of the signal.

The dominant signal in these searches is an observable magnetic field at Earth's surface, which oscillates at the Compton frequency $f_\DM=2\pi/m_\DM$ of the DM~\cite{Fedderke:2021rys,Arza_2022}.%
\footnote{The DPDM signal is, in fact, a sum of three monochromatic signals at frequencies $f=f_{A'},f_{A'}\pm f_d$, where $f_d=1/(\textrm{sidereal day})$ [see \eqref{DPDM_signal}].}
This signal is spatially coherent across the entire globe; i.e., it exhibits a particular spatial pattern so that measurements made at different locations will be correlated.  It is also highly temporally coherent; i.e., it acts as an exact sinusoidal signal\footnotemark[\value{footnote}] for a long coherence time $T_\mathrm{coh}\sim2\pi/(m_\DM v_\DM^2)\sim10^6/f_\DM$, where $v_\DM\sim10^{-3}$ is the typical velocity of the DM.  Equivalently, the signal exhibits a very narrow linewidth $\Delta f\sim10^{-6}f_\DM$ in frequency space.  These properties make the signal ideal to search for in large geomagnetic field datasets, containing data from many geographically distributed stations over long periods of time.

\citeR[s]{Fedderke:2021iqw,Arza_2022} searched for these signals of DPDM and axion DM in the ``low-fidelity" geomagnetic field dataset maintained by the SuperMAG collaboration~\cite{SuperMAGwebsite,Gjerloev:2009wsd,Gjerloev:2012sdg}.  This dataset consists of time series of three-axis magnetic-field measurements from $\mathcal O(500)$ stations located across the globe.  In total, the dataset extends over a period of 50 years (although most stations do not record data for the entire time period).  Importantly, this dataset exhibits a one-minute time resolution, so that the DM searches were only sensitive to Compton frequencies $f_\DM\leq1/(1\,\mathrm{min})$ [corresponding to masses $m_\DM\leq7\times10^{-17}\eV$].  Both the DPDM and axion searches were able to set constraints on parameter space for masses below this cutoff.  The DPDM search of the SuperMAG low-fidelity dataset is the most sensitive direct probe of DPDM in this mass range, but is slightly weaker than existing (indirect) astrophysical probes~\cite{McDermott_2020,Wadekar_2021}.  The axion constraint, on the other hand, is comparable to the constraint from the CAST helioscope~\cite{Anastassopoulos:2017ftl}.

In this work, we extend the analyses of \citeR[s]{Fedderke:2021iqw,Arza_2022} to the SuperMAG ``high-fidelity" dataset, which exhibits a shorter one-second time resolution, albeit with fewer, $\mathcal O(200)$, total stations.  This will enable sensitivity to Compton frequencies up to $f_\DM\leq1\Hz$.  As ambient geomagnetic field noise is typically reduced at higher frequencies, this DM search strategy should be even more powerful at these higher frequencies.  In fact, the SNIPE Hunt collaboration recently conducted an independent search for this same DM signal at even higher frequencies ($0.5\Hz\leq f_\DM\leq5\Hz$) and was able to achieve competitive results with data from only three magnetically quiet locations~\cite{Sulai_2023,Bloch_2024}.  The search in this work leverages the benefits of both low noise at higher frequencies and the large amounts of data that the high-fidelity dataset has to offer.  In this work, we search for ultralight DM in the frequency range $10^{-3}\Hz\leq f_\DM\leq0.98\Hz$; which complements the existing searches from the SuperMAG low-fidelity dataset and the SNIPE Hunt collaboration.

This work is organized as follows.  In \secref{signal}, we briefly review the magnetic-field signal of ultralight DM that we search for.  In \secref{dataset}, we introduce the SuperMAG high-fidelity dataset and review the data-processing techniques that have been applied to it.  In \secref{analysis}, we describe the details of our analysis procedure.  The analysis in this work largely follows the same procedure as in \citeR[s]{Fedderke:2021iqw,Arza_2022} but with important modifications that we highlight in this section.  In \secref{conclusion}, we present and discuss the resulting constraints on ultralight DM from our search.  In \appref{VSH}, we review the properties of vector spherical harmonics (VSH).  In \appref{noise}, we validate our choice of analysis parameters by examining the statistics of the noise in our data.  All the code used in this work is publicly available on Github~\cite{github}.

\section{Signal}
\label{sec:signal}

In \citeR{Fedderke:2021rys}, it was shown that if the DM is comprised of a kinetically mixed dark photon with mass%
\footnote{Throughout this work, in contexts relevant to DPDM, we will write $m_{A'}$ or $f_{A'}$.  In contexts relevant to axion DM, we will write $m_a$ or $f_a$.  In contexts relevant to both, we will write $m_\DM$ or $f_\DM$.}
$m_{A'}=2\pi f_{A'}$ and kinetic mixing parameter $\varepsilon$, it will induce a magnetic-field signal at Earth's surface
\begin{align}
    \bm B_{A'}(\Omega,t)&=\sqrt{\frac\pi3} \left(\varepsilon m_{A'}\right) \left( m_{A'} R \right)\nl
    \cdot\RE\left[ \sum_{m=-1}^1A'_m\bm{\Phi}_{1m}(\Omega)e^{-2\pi i(f_{A'}- f_d m)t} \right].
    \label{eq:DPDM_signal}
\end{align}
Here $\Omega=(\theta,\phi)$ denotes the geographic coordinates on Earth's surface, oriented so that $\theta=0$ corresponds to the Geographic North Pole and $\theta=\pi$ to the Geographic South Pole.  (Note that while $\phi$ coincides with the usual definition of longitude, $\theta$ is not the usual latitude.)  Additionally, $R$ is the radius of Earth, $A'_m$ are complex amplitudes corresponding to the polarization modes of the dark photon,%
\footnote{\label{ftnt:cartesian}%
In our conventions, these amplitudes relate to the Cartesian components of $\bm A'$ via
\begin{equation*}
    A'_x = -\frac 1{\sqrt2} \left( A'_+ - A'_- \right), \quad
    A'_y = -\frac i{\sqrt2} \left(A'_+ + A'_- \right), \quad
    A'_z = A'_0.
\end{equation*}}
and $\bm\Phi_{\ell m}$ are VSH (see \appref{VSH}).  The presence of $f_d=1/(\textrm{sidereal day})$ in the time dependence of the signal arises due to the rotation of Earth.  Here the dark-photon amplitudes $A'_m$ are defined in the inertial celestial frame, while the signal in \eqref{DPDM_signal} is measured in the rotating Earth-fixed frame.  Translating between these frames induces a polarization-dependent shift to the frequency of the signal.  (In frequency space, this will introduce sidebands at $f=f_{A'}\pm f_d$ to the signal.)

\eqref{DPDM_signal} accurately describes the DPDM signal within a coherence time $T_\mathrm{coh}\sim10^6/f_{A'}$.  On longer timescales, the amplitudes $A'_m$ vary stochastically from one coherence time to the next,%
\footnote{In this work, we assume that these amplitudes vary independently, so that the DPDM polarization is randomized from one coherence time to the next.  We note that some production mechanisms (e.g., misalignment) may give rise to DPDM whose polarization does not vary signficantly between coherence times~\cite{Arias:2012az,Caputo_2021}.  In this case, the amplitudes $A'_m$ will maintain the same relative size and phase from one coherence time to the next, but their overall size and phase will still vary.}
with typical size determined by
\begin{equation}\label{eq:DPDM_variance}
    \frac12m_{A'}^2\sum_{m=-1}^1\langle|A'_m|^2\rangle=\rho_\DM,
\end{equation}
where $\rho_\DM\approx0.3\,\mathrm{GeV/cm}^3$ is the local DM density~\cite{Evans:2018bqy}, and $\langle\cdot\rangle$ denotes a time average over many coherence times.  In this work, we will consider the amplitudes $A'_m$ to be drawn from a (isotropic) multivariate Gaussian distribution with variance dictated by \eqref{DPDM_variance}.

In the case that the DM is comprised of axionlike particles, a similar signal arises at Earth's surface~\cite{Arza_2022}.  One crucial difference from the DPDM case is that axion DM requires the presence of a background magnetic field.  In the case of the Earth-transducer effect, this role is played by the static geomagnetic field $\bm B_\oplus$.  Earth's geomagnetic field can be parametrized via the IGRF-13 model~\cite{IGRF}, which provides ``Gauss coefficients" $g_{\ell m}$ and $h_{\ell m}$ for a multipole expansion
\begin{equation}\label{eq:IGRF}
    \bm B_\oplus=\sum_{\ell,m}C_{\ell m} \left(\frac Rr\right)^{\ell+2}\left[(\ell+1)\bm Y_{\ell m}-\bm\Psi_{\ell m}\right],
\end{equation}
where
\begin{align}
    C_{\ell m}=(-1)^m\sqrt{\frac{4\pi(2-\delta_0^m)}{2\ell+1}}\frac{g_{\ell m}-ih_{\ell m}}2.
\end{align}
Since Earth's field slowly drifts over time, the Gauss coefficients $g_{\ell m}$ and $h_{\ell m}$ are time dependent.  The IGRF-13 model provides values for these coefficients at five-year intervals from 1900 to 2020, along with their current time derivative.  These values can be interpolated to find the values of the Gauss coefficients at any time between the beginning of 1900 and the beginning of 2020, and the time derivatives can be used to extrapolate their values up to the end of 2024.

With the static geomagnetic field parametrized in this way, the signal of axion DM with mass $m_a=2\pi f_a$ and coupling to photons $g_{a\gamma}$ is given by~\cite{Arza_2022}
\begin{align}
    \bm B_a(\Omega,t)=g_{a\gamma}(m_a R)&\cdot\IM\left[a_0e^{-2\pi if_at}\right]\nonumber\\
    &\cdot\sum_{\ell,m}\frac{C_{\ell m}(t)}\ell\bm\Phi_{\ell m}(\Omega).
    \label{eq:axion_signal}
\end{align}
Unlike the DPDM case, axion DM is described by only a single amplitude $a_0$.  Again within a coherence time $T_\mathrm{coh}\sim10^6/f_a$, this amplitude is constant and \eqref{axion_signal} remains valid; on longer timescales, however, $a_0$ varies stochastically with variance given by
\begin{equation}
    \frac12m_a^2\langle|a_0|^2\rangle=\rho_\DM.
\end{equation}

Before proceeding, we note a few distinctions between the signals in \eqref[s]{DPDM_signal} and (\ref{eq:axion_signal}).  First, the DPDM signal is purely dipolar (only $\ell=1$ modes contribute).  Because $\bm B_\oplus$ is approximately dipolar, the expansion in \eqref{IGRF} is dominated by the $C_{10}$ term, but other terms contribute at the $\mathcal O(10\%)$ level.  Likewise, the axion signal in \eqref{axion_signal} is predominantly dipolar, but receives small contributions from higher $\ell$ modes.  Secondly, the axion signal in \eqref{axion_signal} does not feature any dependence on $f_d$.  This is because the angular dependence of the axion signal is set by $\bm B_\oplus$, which corotates with Earth, as opposed to the DPDM polarization which does not.  This means that, in frequency space, the axion signal will only appear at $f=f_a$, whereas the DPDM signal will appear at $f=f_{A'},f_{A'}\pm f_d$.  Finally, the global angular dependence of the DPDM signal will vary from one coherence time to the next, as the amplitudes $A'_m$ stochastically vary.  The global angular dependence of the axion signal, however, does not change as $a_0$ stochastically varies.  Instead, the axion signal will slowly drift along with $\bm B_\oplus$.  We note that $C_{10}$ only changes by $\mathcal O(1\%)$ over the 23-year duration of the high-fidelity dataset, so this is a much slower timescale.

\section{SuperMAG high-fidelity dataset}
\label{sec:dataset}

The SuperMAG collaboration maintains a large database of three-axis magnetometer measurements from stations dispersed across the globe, which it makes publicly available for research purposes~\cite{SuperMAGwebsite,Gjerloev:2009wsd,Gjerloev:2012sdg}.  In this work, we search their ``high-fidelity" dataset which reports magnetic-field measurements at one-second intervals.  This dataset contains many fewer stations than their ``low-fidelity" (one-minute time resolution) dataset, which was analyzed in \citeR[s]{Fedderke:2021iqw,Arza_2022}, as not all stations report higher-frequency data.  Likewise it spans a shorter period of time than the low-fidelity dataset.  Many of the features of the high-fidelity dataset and the data-processing techniques applied to it are similar to those of the low-fidelity dataset, which were outlined in \citeR{Fedderke:2021iqw}.  Here we will briefly review the important features, but we refer the interested reader to that prior work for a more detailed summary, or to \citeR{Gjerloev:2012sdg} for a thorough description of the SuperMAG collaboration's data-processing techniques.

\begin{figure}[!t]
\includegraphics[width=\columnwidth]{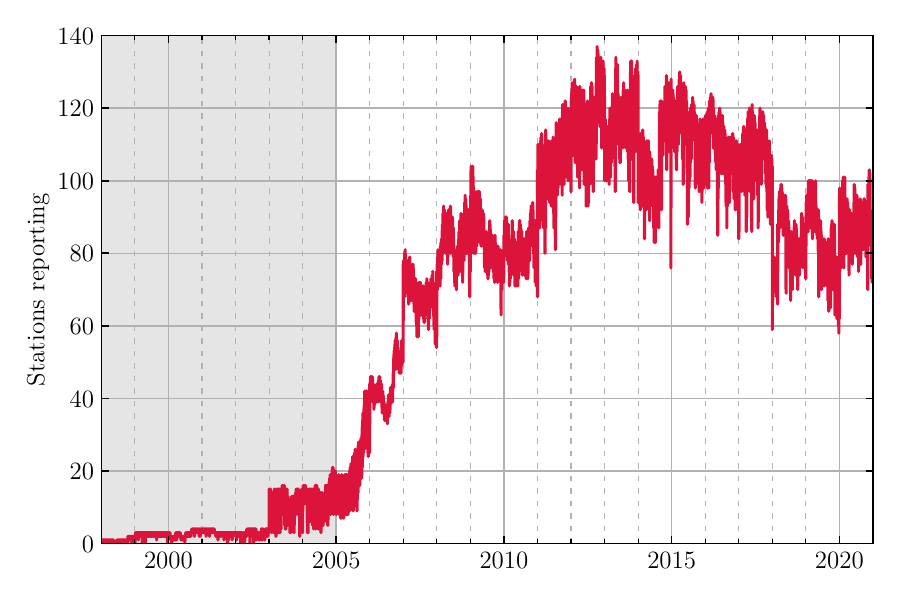}
\caption{\label{fig:count}%
    Number of stations in the SuperMAG dataset~\cite{SuperMAGwebsite,Gjerloev:2009wsd,Gjerloev:2012sdg} that are actively reporting, as a function of the date.  The shaded gray region indicates the data which are not utilized in this work due to an insufficient number of stations and/or the presence of transient features.}
\end{figure}

The high-fidelity dataset analyzed in this work covers the period from the beginning of 1998 to the end of 2020.  In total, the dataset combines measurements from 246 stations, but most stations only actively report data for a small portion of this full duration.  In \figref{count}, we show the number of actively reporting stations over time, which peaks in 2012 with $\sim130$ active stations.  Prior to the year 2003, the dataset only contains measurements from four stations.  For technical reasons, our DPDM analysis requires at least three active stations at any given time,%
\footnote{In our DPDM analysis, we combine the north and east magnetic-field measurements from all the active stations to form five time series $X^{(1)},\ldots,X^{(5)}$ which we then search for a DPDM signal.  We require at least three stations (six magnetic-field measurements total) in order for these time series to be linearly independent.}
and so this region of the dataset is unsuitable for our analysis.  In addition, the data from 2003 and 2004 were found to contain extended periods of excessive noise.  For these reasons, the analysis in this work focuses on the portion of the dataset beginning in 2005.  Since few stations were active before 2005, we expect that discarding this data will not significantly affect our final sensitivity to ultralight DM.

The three-axis magnetic-field data measured at each station are reported by SuperMAG in coordinates aligned with the local magnetic field; that is, each station reports the components $B_\mathrm{LMN}$ and $B_\mathrm{LME}$ of the magnetic field along Local Magnetic North and Local Magnetic East, respectively, along with the vertical component $B_\mathrm{Z}$.  These local magnetic-field coordinates are defined so that the ``typical" value of $B_\mathrm{LME}$ is zero over a 17-day sliding window~\cite{Gjerloev:2012sdg}.  SuperMAG also provides time-dependent declination angles $\delta(\Omega_i,t)$ for each station which can be used to convert to coordinates aligned with True Geographic North and True Geographic East
\begin{equation}
    \begin{pmatrix}B_\mathrm{TGN}\\B_\mathrm{TGE}\end{pmatrix}=\begin{pmatrix}\cos\delta&-\sin\delta\\\sin\delta&\cos\delta\end{pmatrix}\begin{pmatrix}B_\mathrm{LMN}\\B_\mathrm{LME}\end{pmatrix}.
\end{equation}
The spherical coordinates we utilize in this work are related to these geographic coordinates via
\begin{align}
    B^\theta&=-B_\mathrm{TGN}\\
    B^\phi&=B_\mathrm{TGE}\\
    B^r&=-B_\mathrm{Z}.
\end{align}
(SuperMAG's conventions are such that a positive $B_\mathrm{Z}$ indicates a vertically downward-pointing magnetic field, where we define $\rhat$ to point upwards.)

SuperMAG applies a rigorous data-cleaning procedure to the raw data it receives from its collaborators (prior to rotating coordinates).  This procedure consists of three steps: automatic removal of data spikes; manual removal of errors; and automatic correction for instrument offsets/biases.  First, spikes of duration five seconds or less are identified, removed, and replaced by linear interpolation. Next, manual identification and correction of other data errors are performed by highly experienced personnel.  Last, instruments typically exhibit time-dependent offsets in each component that are corrected using known main-field models.  Any temperature-dependent gain factors exhibited by the magnetometers are corrected for as well.

The dataset utilized in this work has also had a yearly and a daily baseline subtracted to remove any manifestly annual or diurnal features in the data~\cite{Gjerloev:2012sdg}.  As noted in \citeR{Fedderke:2021iqw}, these timescales (as well as the 17-day and 30-minute windows utilized in the yearly and daily baseline removal procedures, respectively) are well below the frequency range analyzed in this work, and so we do not expect them to affect the conclusions of our analysis.

Finally, we note that we have explicitly removed a few obvious anomalies from the dataset, specifically for the purpose of this analysis.  The reported measurements in the dataset exceed $10^5\,\mathrm{nT}$ in six instances, in each case only for one or two seconds and immediately preceding a period of inactivity.  Due to the anomalously large size and short duration of these measurements, we believe them to be artifacts of the post-processing procedure.  We replace all such measurements with null values (as if the station were inactive).  In addition, the data reported by the single station located at Lucky Lake (IAGA code:~LCL) exhibit a number of single-second spikes.  These spikes have low amplitude, so that our analysis would actually upweight this station's measurements (see discussion around \eqref[s]{weights_theta} and (\ref{eq:weights_phi}) for weighting procedure based on white noise level).  This station is therefore unsuitable for common-noise modeling treatment with the other stations.  Due to the prevalence and duration of these spikes, we do not suspect them to be physical magnetic-field measurements.  In lieu of attempting to remedy these anomalies in post-processing, we simply discard the data from this single station.

\section{Analysis details}
\label{sec:analysis}

In this section, we describe the analysis of the high-fidelity dataset.  This analysis is based on the analyses performed in \citeR[s]{Fedderke:2021iqw,Arza_2022} on the low-fidelity dataset.  The analyses in those works were complicated by the fact that the SuperMAG dataset consists of data from many stations that are active at different times.  The strategy in \citeR[s]{Fedderke:2021iqw,Arza_2022} was to first combine the time series of measurements from all $\mathcal O(500)$ stations into a handful of specific linear combinations whose forms were motivated by \eqref[s]{DPDM_signal} and (\ref{eq:axion_signal}).  \citeR[s]{Fedderke:2021iqw,Arza_2022} then partitioned these constructed time series into chunks of length roughly equal to the coherence time $T_\mathrm{coh}$ [see discussion around \eqref{DPDM_variance}] and searched for the DM signals in each coherence chunk independently.  In order to perform this search, it was important to both understand how a potential signal would appear in these constructed time series and to characterize the noise levels of these time series.  The former was done by applying the same combination procedure that was used on the magnetometer data to the signals in \eqref[s]{DPDM_signal} and (\ref{eq:axion_signal}), while the latter was estimated via a data-driven procedure.  Once the expected signal size and noise levels were understood, \citeR[s]{Fedderke:2021iqw,Arza_2022} computed a posterior distribution on the coupling $\varepsilon$ or $g_{a\gamma}$, which was then used to either set constraints on these parameters or identify signal candidates.

In this work, we follow the same strategy as in \citeR[s]{Fedderke:2021iqw,Arza_2022}, but make the following modifications:
\begin{itemize}
    \item\emph{Stationarity timescale}: The analysis in \citeR[s]{Fedderke:2021iqw,Arza_2022}
    relied on the assumption that the noise in the constructed time series was stationary within a calendar year.  To this end, the weights used in constructing the time series were chosen on a calendar-year basis, and noise spectra were computed for each calendar year separately.  In \appref{stationarity}, we re-evaluate what constitutes an appropriate choice of ``stationarity timescale" for our analysis of the high-fidelity dataset.  In this work, we take this timescale to be a week instead of a calendar year.%
    \footnote{As the first year of data we utilize is 2005, the first stationarity period is taken to start on Saturday, January 1, 2005.  Therefore, the stationarity periods are always Saturday through Friday.  The final stationarity period is extended to contain the remainder of the dataset, so it is actually 13 days long.}
    \item\emph{Noise estimation}: Our analysis requires a procedure for estimating the noise levels in the constructed time series.  (As stated above, the noise is estimated separately for each ``stationarity period" of a week.)  In this work, we adopt a slightly simpler approach than the one in \citeR[s]{Fedderke:2021iqw,Arza_2022}.  For each stationarity period, we simply compute the power spectral density (PSD) of the time series over the full period and smooth it with a Gaussian filter to obtain a noise spectrum.  (See \appref{window} for choice of filter window size.)
    \item\emph{Candidate validation}: The analyses in \citeR[s]{Fedderke:2021iqw,Arza_2022} each identified $\mathcal O(30)$ potential DM candidates.  In order to further scrutinize whether these candidates exhibited properties of a true DM candidate, \citeR[s]{Fedderke:2021iqw,Arza_2022} developed a candidate validation procedure, which repeated the analyses on various subsets of the data (either a subset of stations or subset of temporal duration) to see if the candidate persisted.  Ultimately, this procedure was able to definitively reject most of the DM candidates, and the remaining were argued to be weak candidates.  In this work, we instead find $\mathcal O(1000)$ potential candidates, and so we require a much stronger protocol to effectively scrutinize all of them.  We develop a refined candidate validation procedure in \secref{candidates}, which is able to reject all candidates as robust DM signals.
\end{itemize}
We now describe the analysis procedure in detail, paying special attention to these new features that differ from \citeR[s]{Fedderke:2021iqw,Arza_2022}.

\subsection{Time series}
\label{sec:timeseries}

Each SuperMAG station $i$, located at coordinates $\Omega_i=(\theta_i,\phi_i)$, reports a time series of three-axis magnetic-field measurements $\bm B(\Omega_i,t_j)$ for some set of sampling times $t_j\in\mathcal T_i$.  The set $\mathcal T_i$ of sampling times at which a particular station is active differs from one station to the next, making the prospect of analyzing stations separately and combining their results challenging.  Instead, as in \citeR[s]{Fedderke:2021iqw,Arza_2022}, we first combine the time series of measurements from the individual stations into a handful of specific linear combinations and then analyze these constructed time series as our input.

The manner of combining the time series will be motivated by the signals in \eqref[s]{DPDM_signal} and (\ref{eq:axion_signal}).  That is, in the DPDM case, we project the magnetic-field measurements from each station onto the real and imaginary parts of the components of the $\bm\Phi_{1m}$ VSH
\begin{align}
    X_i^{(1)}(t_j) &\equiv \sin\phi_i \cdot B_i^\theta(t_j), \label{eq:X1i} \\
    X_i^{(2)}(t_j) &\equiv \cos\phi_i \cdot B_i^\theta(t_j), \label{eq:X2i}\\
    X_i^{(3)}(t_j) &\equiv \cos\phi_i\cos\theta_i \cdot B_i^\phi(t_j), \label{eq:X3i} \\
    X_i^{(4)}(t_j) &\equiv -\sin\phi_i\cos\theta_i \cdot B_i^\phi(t_j), \label{eq:X4i}\\
    X_i^{(5)}(t_j) &\equiv \sin\theta_i \cdot B_i^\phi(t_j). \label{eq:X5i}
\end{align}
On the other hand, in the axion case, we project them onto the components of the linear combination in \eqref{axion_signal}
\begin{align}\label{eq:Xtheta}
    X_i^{(1)}(t_j) &= \left(\sum_{\ell,m}\frac{C_{\ell m}(t_j)}\ell\Phi_{\ell m}^\theta(\Omega_i)\right)\cdot B_i^\theta(t_j),\\
    X_i^{(2)}(t_j) &= \left(\sum_{\ell,m}\frac{C_{\ell m}(t_j)}\ell\Phi_{\ell m}^\phi(\Omega_i)\right)\cdot B_i^\phi(t_j).
\label{eq:Xphi}\end{align}

We can then define time series by a weighted combination of the projections from all the stations
\begin{equation}\label{eq:Xn}
    X^{(n)}(t_j) = \frac1{W^{(n)}(t_j)}\sum_{\{i|t_j\in\mathcal T_i\}}w_i^{(n)}(t_j) X_i^{(n)}(t_j),
\end{equation}
The sum here is over all stations $i$ that are active at time $t_j$.  As in \citeR[s]{Fedderke:2021iqw,Arza_2022}, the station weights $w_i^{(n)}(t_j)$ should be constant within a stationarity period (a week in the case of this work's analysis) but may vary from one period to the next.  In order to downweight noisier stations, we set the weights to be the inverse of the white noise level of the station; that is, for all $t$ within stationarity period $a$, we take
\begin{align}
    \label{eq:weights_theta}
    w_i^{\theta}(t)&=\left(\frac1{N_i^a}\sum_{t_j\in\mathcal T_i^a} \left[B_i^\theta(t_j)\right]^2\right)^{-1}, \\
    w_i^{\phi}(t)&=\left(\frac1{N_i^a}\sum_{t_j\in\mathcal T_i^a}\left[B_i^\phi(t_j)\right]^2\right)^{-1},
    \label{eq:weights_phi}
\end{align}
where $\mathcal T_i^a$ is the subset of $\mathcal T_i$ within stationarity period $a$ and $N_i^a$ is the number of sampling times in $\mathcal T_i^a$.  In the DPDM case, we set $w_i^{(1)}=w_i^{(2)}=w_i^\theta$ and $w_i^{(3)}=w_i^{(4)}=w_i^{(5)}=w_i^\phi$.  In the axion case, we set $w_i^{(1)}=w_i^\theta$ and $w_i^{(2)}=w_i^\phi$.  The total weights $W^{(n)}(t_j)$ appearing in \eqref{Xn} are the sum of the active station weights at a given time $t_j$; that is
\begin{equation}
    W^{(n)}(t_j)=\sum_{\{i|t_j\in\mathcal T_i\}}w_i^{(n)}(t_j).
\end{equation}
Note that, although the station weights $w_i^{(n)}$ are constant within a stationarity period, the total weights may not be because the number of active stations over which this sum runs may vary within a stationarity period.

\subsection{Data vector}
\label{sec:datavector}

Once the time series $X_i^{(n)}$ have been constructed, we will deal solely with these projected variables rather than the data from individual stations.  Our goal is to search these time series for a nearly monochromatic signal.  A natural strategy would therefore be to Fourier transform the data and look for excess power at a particular frequency.  However, as described in \secref{signal}, the signal can only be treated as monochromatic within a coherence time $T_\mathrm{coh}\sim10^6/f_\DM$.  Therefore, we will instead partition the time series into ``coherence chunks" each roughly of length $T_\mathrm{coh}$ and search each chunk independently for a monochromatic signal.  We will then combine the results of these searches incoherently (i.e., we will treat them as independent searches).

Because the coherence time $T_\mathrm{coh}$ depends explicitly on the DM frequency, this strategy should, in principle, require a different partition of the time series for each DM frequency we wish to search for.  Such a procedure would be computationally intractable.  Instead, \citeR{Fedderke:2021iqw} developed a procedure for choosing an \textit{approximate}%
\footnote{In \citeR{Fedderke:2021iqw} and in this work, this approximation of the coherence time holds to 3\%.  We note that the notion of a coherence time is itself approximate, as it is determined by $v_\DM$, which is known even less precisely than this.  Therefore the use of this approximation (and the choice of a 3\% tolerance) introduces no more error into our analysis than our use of the coherence time in the first place.}
coherence time that could apply to a range of DM frequencies.  In this way, the entire set of DM frequencies to be searched for could be split into $\mathcal O(50)$ ranges, each range could be assigned an approximate coherence time (which would differ from the true $T_\mathrm{coh}$ by at most 3\% for all frequencies in the range), and the same partition of the time series could be used for the entire range.  In this work, we apply the same procedure for selecting the DM frequencies to be searched for and determining the approximate coherence times.  We restrict our analysis to $10^{-3}\Hz\leq f_\DM\leq0.98\Hz$,%
\footnote{We choose this upper bound because our analysis of the noise statistics in \appref{window} computes these statistics by averaging (in log-space) over a window of size $\sigma_{\log}=0.02$.  We therefore cannot verify our assumptions on the noise statistics at frequencies $f_\DM\gtrsim0.98\Hz$.}
which requires $106$ frequency ranges.  Rather than reiterating the details of this procedure, we refer the interested reader to Sec.~V\,E of \citeR{Fedderke:2021iqw}.  Henceforth, we will focus on the search for a single DM frequency $f_\DM$; we will simply refer to its approximate coherence time as $T_\mathrm{coh}$.

Once we have fixed the coherence time $T_\mathrm{coh}$, we partition the full time series $X^{(n)}(t)$ into coherence chunks $X^{(n)}_k(t)$ each of length $T_\mathrm{coh}$.  For each $k$, we wish to combine all the information in the time series $X^{(n)}_k$ that is relevant to searching for a DM signal of frequency $f_\DM$ into a single vector.  Note that in the DPDM case, the signal may contain information at the frequency $f=f_{A'}$, as well as $f=f_{A'}\pm f_d$.  Therefore, our data vector should include the Fourier transforms of all five time series at all three of these frequencies, that is we should define the 15-dimensional vector%
\footnote{As $f_d$ may not generically be a discrete Fourier transform (DFT) frequency, it may be computationally difficult to compute these Fourier transforms at $f=f_{A'}\pm f_d$.  (The prescription for choosing frequencies outlined in Sec.~V\,E of \citeR{Fedderke:2021iqw} is designed so that $f_{A'}$ is always a DFT frequency.)  As a result, we instead evaluate them at $f=f_{A'}\pm\hat f_d$, where $\hat f_d$ is the DFT frequency closest to $f_d$.}%
\textsuperscript{,}%
\footnote{Throughout this section, $\vec x$ denotes a 15-dimensional vector in the DPDM case and a two-dimensional vector in the axion case.  Meanwhile, $\bm y$ denotes a vector with three (typically spatial) dimensions.}
\begin{equation}
    \vec X_k = \begin{pmatrix}
	\tilde X_k^{(1)}(f_{A'}-\hat f_d)\\
	\tilde X_k^{(2)}(f_{A'}-\hat f_d)\\
	\tilde X_k^{(3)}(f_{A'}-\hat f_d)\\
	\tilde X_k^{(4)}(f_{A'}-\hat f_d)\\
	\tilde X_k^{(5)}(f_{A'}-\hat f_d)\\
	\tilde X_k^{(1)}(f_{A'})\\
	\tilde X_k^{(2)}(f_{A'})\\
	\tilde X_k^{(3)}(f_{A'})\\
	\tilde X_k^{(4)}(f_{A'})\\
	\tilde X_k^{(5)}(f_{A'})\\
	\tilde X_k^{(1)}(f_{A'}+\hat f_d)\\
	\tilde X_k^{(2)}(f_{A'}+\hat f_d)\\
	\tilde X_k^{(3)}(f_{A'}+\hat f_d)\\
	\tilde X_k^{(4)}(f_{A'}+\hat f_d)\\
	\tilde X_k^{(5)}(f_{A'}+\hat f_d)
    \end{pmatrix},
\end{equation}
where $\tilde X_k^{(n)}(f)$ denotes the Fourier transform of $X_k^{(n)}(t)$.  In the axion case, the only relevant frequency is $f=f_a$, and so we instead simply define a two-dimensional data vector
\begin{equation}
    \vec X_k = \begin{pmatrix}
        \tilde X_k^{(1)}(f_a)\\
	\tilde X_k^{(2)}(f_a)
    \end{pmatrix}.
\end{equation}

These data vectors will be the primary objects from which we will construct a likelihood analysis.  In order to construct a likelihood for them, it is important to understand both their expectation (in the presence of a signal) and their covariance.  To compute the expectation, one substitutes the signals in \eqref[s]{DPDM_signal} and (\ref{eq:axion_signal}) into the definitions in \eqrefRange{X1i}{X5i} and \eqrefRange{Xtheta}{Xphi}, respectively, and follows through the process of constructing the data vector.  In the DPDM case, this yields an expression of the form
\begin{equation}
    \langle\vec X_k\rangle=\varepsilon(c_{xk}^*\vec\mu_{xk}+c_{yk}^*\vec\mu_{yk}+c_{zk}^*\vec\mu_{zk}),
\end{equation}
where $c_{ik}$ for $i=x,y,z$ are related to the Cartesian components of $\bm A'_k$ (the polarization of $\bm A'$ within the $k$-th coherence chunk) by
\begin{equation}\label{eq:DPDM_ck}
    c_{ik}\equiv\frac{\sqrt2\pi f_{A'}}{\sqrt{\rho_\DM}}A'_{ik}.
\end{equation}
(See footnote~\ref{ftnt:cartesian} for the relationship between the Cartesian components $A'_i$ and polarization amplitudes $A'_m$.)  These are normalized so that \eqref{DPDM_variance} becomes $\langle|\bm c_k|^2\rangle=1$.  We refer the reader to Eqs.~(36), (C1), and (C2) of \citeR{Fedderke:2021iqw} for the explicit expressions for $\vec\mu_{ik}$.  In the axion case, the expectation value of $\vec X_k$ (in the presence of a signal) is
\begin{align}
    \langle\vec X_k\rangle&=g_{a\gamma}c^*_k\vec\mu_k,\\
    c_k&\equiv\frac{\sqrt2\pi f_a}{\sqrt{\rho_\DM}}a_{0k},
\label{eq:axion_ck}\end{align}
where $a_{0k}$ is the amplitude $a_0$ within the $k$-th coherence chunk.  The explicit expression for $\vec\mu_k$ in the axion case is given by Eq.~(C10) of \citeR{Arza_2022}.

\subsection{Noise estimation}
\label{sec:noise}

Now that we have computed the expectation of $\vec X_k$, we must also compute its variance.  
Doing so requires an estimate of the noise level in the time series.  As our weights were chosen in \secref{timeseries} to be constant within a stationarity period, we will henceforth assume that the noise level within a stationarity period is roughly constant.  Our goal is then to determine the cross-power spectral density $S_{mn}^a$ between the noise in time series $X_a^{(m)}(t_j)$ and $X_a^{(n)}(t_j)$ within a given stationarity period $a$.  If $x_a^{(n)}(t_j)$ represents a noise-only realization of the time series for a duration $\tau$ entirely contained within stationarity period $a$, and $\tilde x_a^{(n)}(f_p)$ represents its discrete Fourier transform (DFT) evaluated at DFT frequencies $f_p=p/\tau$, then the two-sided cross-power spectral density of the noise in stationarity period $a$ is defined by
\begin{equation}\label{eq:noisePSD}
    \langle\tilde x_a^{(m)}(f_p)\, \tilde x_a^{(n)}(f_q)^*\rangle_{\varepsilon=0} \equiv \tau S_{mn}^a(f_p)\,\delta_{pq},
\end{equation}
where $\langle\,\cdots\rangle_{\varepsilon=0}$ indicates the expectation over all noise realizations (i.e., with no signal, $\varepsilon=0$).

In \citeR[s]{Fedderke:2021iqw,Arza_2022}, the noise PSD $S_{mn}^a$ was estimated via a data-driven procedure, in which the full stationarity period was divided into chunks.  Each chunk was then treated as an independent realization of the noise, so that averaging over the chunks would simulate the expectation in \eqref{noisePSD}.  This method is valid so long as each chunk is statistically independent from the next.  In our analysis of the high-fidelity dataset, we observe small correlations between such chunks, and so we opt for a slightly different noise-estimation procedure in this work.

Rather than simulating the expectation in \eqref{noisePSD} by an average in time (i.e., over temporally separated chunks), we simulate it by an average in frequency.  That is, we assume that $S_{mn}^a(f)$ is roughly constant over a sufficiently small frequency range.  Therefore by averaging the power over several neighboring frequency bins, we can obtain a good estimate of $S_{mn}^a(f_p)$.  Ultimately, this simply amounts to computing the PSD of $X_a^{(n)}$ over the full stationarity period and smoothing it by averaging over neighboring frequency bins.  In this work, we utilize a Gaussian filter to perform this smoothing, that is, we estimate the noise PSD as
\begin{align}\label{eq:PSDsmoothing}
    S_{mn}^a(f_p)&=\frac1{T^a}\sum_q\mathcal K(q-p;\sigma)\tilde X_a^{(m)}(f_q)\tilde X_a^{(n)}(f_q)^*,\\
    \mathcal K(p;\sigma)&\equiv\frac1{\sigma\sqrt{2\pi}}\exp\left(-\frac{p^2}{2\sigma^2}\right),
\label{eq:filter}\end{align}
where $T^a$ is the duration of stationarity period $a$ (usually one week), and $\sigma=15$ frequency bins is the window size for the Gaussian filter (see \appref{window} for validation of this choice).  \eqref{PSDsmoothing} allows us to compute the noise PSD at any DFT frequency $f_p=p/T^a$.  As $f_\DM$ may not generically be one of these frequencies, we typically must interpolate from these DFT frequencies to $f_\DM$ (and $f_{A'}\pm\hat f_d$ in the DPDM case).

Once the noise PSD for each stationarity period has been computed, the total variance of $\tilde X_k^{(n)}(f)$ can be determined by combining the noise PSDs from all the stationarity periods that the coherence chunk $k$ overlaps with, that is
\begin{equation}\label{eq:Xk_variance}
    \langle\tilde X_k^{(m)}(f)\tilde X_k^{(n)}(f)^*\rangle_{\varepsilon=0}=\sum_a T_k^a \cdot S_{mn}^a(f),
\end{equation}
where $T_k^a$ is the duration of the overlap between stationarity period $a$ and coherence chunk $k$.  The covariance matrix of $\vec X_k$ in the DPDM case is then a block diagonal matrix
\begin{widetext}
\begin{equation}\label{eq:DPDM_sigma}
    \Sigma_k\equiv\mathrm{Cov}(\vec X_k,\vec X_k)
    =\begin{pmatrix}\sum_aT_k^a\cdot S_{mn}^a(f_{A'}-\hat f_d)\\&\sum_aT_k^a\cdot S_{mn}^a(f_{A'})\\&&\sum_aT_k^a\cdot S_{mn}^a(f_{A'}+\hat f_d)\end{pmatrix},
\end{equation}
\end{widetext}
while in the axion case it is simply
\begin{equation}\label{eq:axion_sigma}
    \Sigma_k=\sum_a T_k^a \cdot S_{mn}^a(f_a).
\end{equation}

\subsection{Bayesian statistical analysis}
\label{sec:bayesian}

With the expectation and variance of $\vec X_k$ computed, we can proceed with defining a likelihood and calculating a bound.  We model each $\vec X_k$ as a multivariate Gaussian variable (see \appref{window} for validation of their Gaussianity), so that they are entirely characterized by their expectation and variance.  We consider each coherence chunk $k$ as independent, so that the total likelihood is simply the sum over the likelihoods for each $\vec X_k$.  Specifically, the total likelihood is given by
\begin{widetext}
\begin{align}\label{eq:DPDM_likelihood}
    -\ln\LL\left(\varepsilon,\{\bm c_k\}\big|\{\vec X_k\}\right)&=\sum_k\left(\vec X_k-\varepsilon\sum_ic_{ik}^*\vec\mu_{ik}\right)^\dagger\Sigma_k^{-1}\left(\vec X_k-\varepsilon\sum_ic_{ik}^*\vec\mu_{ik}\right),\\
    -\ln\LL\left(g_{a\gamma},\{c_k\}\big|\{\vec X_k\}\right)&=\sum_k\left(\vec X_k-g_{a\gamma}c_k^*\vec\mu_k\right)^\dagger\Sigma_k^{-1}\left(\vec X_k-g_{a\gamma}c_k^*\vec\mu_k\right),
\label{eq:axion_likelihood}\end{align}
\end{widetext}
in the DPDM and axion cases respectively.

Our likelihood analysis can be simplified by making a few changes of variables to \eqref[s]{DPDM_likelihood} and (\ref{eq:axion_likelihood}).  As $\Sigma_k$ is a positive-definite Hermitian matrix, it can be written as $\Sigma_k=A_kA_k^\dagger$ for some invertible $A_k$.  Then we may define
\begin{align}
    \vec Y_k &=A_k^{-1}\vec X_k,&
    \vec\nu_{(i)k}&=A_k^{-1}\vec\mu_{(i)k},
\end{align}
(where the $i$ index is only present in the DPDM case).  In the DPDM case, we combine the three 15-dimensional vectors $\vec\nu_{ik}$ into a single $15\times3$ matrix $N_k$, and take its singular value decomposition
\begin{equation}
    N_k=U_kS_kV_k^\dagger,
\end{equation}
where $U_k$ is a $15\times3$ matrix with orthonormal columns, $S_k$ is a real $3\times3$ diagonal matrix, and $V_k$ is a unitary $3\times3$ matrix.  We then define the three-dimensional vectors
\begin{align}\label{eq:dk}
    \bm d_k&=V_k^\dagger\bm c_k^{*},&
    \bm Z_k&=U_k^\dagger\vec Y_k.
\end{align}
With these variables, \eqref{DPDM_likelihood} can be written as
\begin{equation}\label{eq:DPDM_likelihood_Z}
    -\ln\LL\left(\varepsilon,\{\bm d_k\}\big|\{\bm Z_k\}\right)=\sum_k|\bm Z_k-\varepsilon S_k\bm d_k|^2.
\end{equation}
The analogous change of variables in the axion case is simply
\begin{align}
    s_k&=|\vec\nu_k|,&
    z_k&=\frac{\vec\nu_k^\dagger\vec Y_k}{s_k},
\end{align}
so that \eqref{axion_likelihood} becomes
\begin{equation}\label{eq:axion_likelihood_Z}
    -\ln \LL\left(g_{a\gamma},\{c_k\}\big|\left\{z_k\right\}\right)=\sum_k\left|z_k-g_{a\gamma}c_k^*s_k\right|^2.
\end{equation}

We apply a Bayesian analysis framework in order to convert the likelihoods in \eqref[s]{DPDM_likelihood_Z} and (\ref{eq:axion_likelihood_Z}) into posteriors on the couplings $\varepsilon$ and $g_{a\gamma}$.  We treat the amplitudes $\bm d_k$ and $c_k$ as nuisance parameters, so we must first marginalize over them.  As mentioned in \secref{signal}, we take the amplitudes to have Gaussian distributions.  \eqref[s]{DPDM_ck} and (\ref{eq:axion_ck}) define them to have unit variance (and multiplying by $V_k^\dagger$ does not change their normalization), so their likelihoods are given by
\begin{align}
    \LL\left(\bm d_k\right)&=\exp\left(-3|\bm d_k|^2\right)\\
    \LL\left(c_k\right)&=\exp\left(-|c_k|^2\right).
\end{align}
Marginalizing \eqref[s]{DPDM_likelihood_Z} and (\ref{eq:axion_likelihood_Z}) over these distributions of $\bm d_k$ and $c_k$ then gives likelihoods for only the couplings
\begin{align}
    \LL(\varepsilon\big|\{\bm Z_k\})&\propto\prod_{i,k}\frac1{3+\varepsilon^2s_{ik}^2}\exp\left(-\frac{3|z_{ik}|^2}{3+\varepsilon^2s_{ik}^2}\right),\\
    \LL(g_{a\gamma}\big|\{z_k\})&\propto\prod_k\frac1{1+g_{a\gamma}^2s_k^2}\exp\left(-\frac{|z_k|^2}{1+g_{a\gamma}^2s_k^2}\right),
\end{align}
where $z_{ik}$ are the components of $\bm Z_k$ and $s_{ik}$ are the diagonal entries of $S_k$ (see Appendix D\,1 of \citeR{Fedderke:2021iqw} for derivation).

To turn these likelihoods into posteriors, we require a prior on the couplings.  \citeR[s]{Fedderke:2021iqw,Arza_2022} utilized the Jeffreys prior,%
\footnote{For a discussion on the choice of different priors in ultralight DM searches, we point the interested reader to Appendix I\,A of \citeR{Centers:2019dyn}, which compares a Jeffreys prior to a uniform prior.  They find the choice of prior only results in an $\mathcal O(1)$ rescaling of the constraint.  Moreover, they point out that the Jeffreys prior is conceptually preferable due to its reparametrization invariance.}
which in this context takes the form
\begin{align}
    p(\varepsilon)&\propto\sqrt{\sum_{i,k}\frac{4\varepsilon^2s_{ik}^4}{\left(3+\varepsilon ^2s_{ik}^2\right)^2}},\\
    p(g_{a\gamma})&\propto\sqrt{\sum_k\frac{4g_{a\gamma}^2s_k^4}{\left(1+g_{a\gamma}^2s_k^2\right)^2}}
\end{align}
(see Appendix D\,2 of \citeR{Fedderke:2021iqw} for derivation).  The posterior distributions of $\varepsilon$ and $g_{a\gamma}$ are then defined by the product
\begin{align}
    p(\varepsilon\big|\{\bm Z_k\})&\propto\LL(\varepsilon\big|\{\bm Z_k\})\cdot p(\varepsilon),\\
    p(g_{a\gamma}\big|\{z_k\})&\propto\LL(g_{a\gamma}\big|\{z_k\})\cdot p(g_{a\gamma}),
\end{align}
up to a normalization.  The normalization is fixed by setting the integral of the posterior over all $\varepsilon$ or $g_{a\gamma}$ to 1.  The 95\%-credible upper limit on the coupling is then defined by
\begin{align}
    \int_0^{\hat\varepsilon}d\varepsilon~p(\varepsilon|\{\bm Z_k\})&=0.95,\\
    \int_0^{\hat g_{a\gamma}}dg_{a\gamma}~p(g_{a\gamma}|\{z_k\})&=0.95.
\end{align}

Last, as in \citeR[s]{Fedderke:2021iqw,Arza_2022}, we apply a 25\% degradation factor to our final upper limits
\begin{align}\label{eq:DPDM_degrade}
    \hat\varepsilon&\rightarrow1.25\cdot\hat\varepsilon,\\
    \hat g_{a\gamma}&\rightarrow1.25\cdot\hat g_{a\gamma}
\label{eq:axion_degrade}\end{align}
to account for the finite width of the signal.  The reason for this is that, thus far, we have assumed that the DM signal within a given coherence chunk would be exactly monochromatic.  In fact, due to the approximate nature of the coherence time, a true DM signal would be spread across a few frequency bins, thereby slightly decreasing the power in the one central bin in which we have searched.  In \citeR{Fedderke:2021iqw}, it was determined that a 25\% reduction in the final constraint was sufficient to correct for such a loss of signal power.

\subsection{Candidate identification and validation}
\label{sec:candidates}

\begin{figure*}[t]
\includegraphics[width=0.49\textwidth]{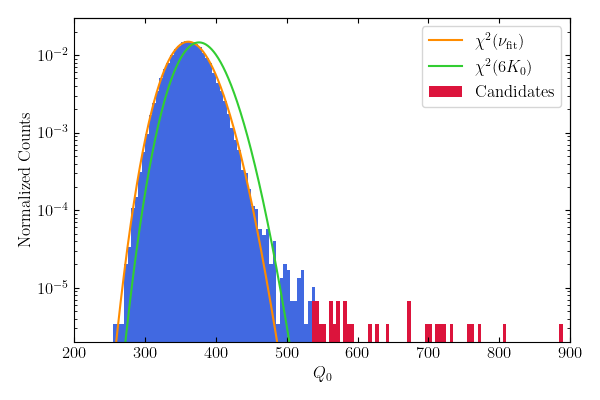}
\includegraphics[width=0.49\textwidth]{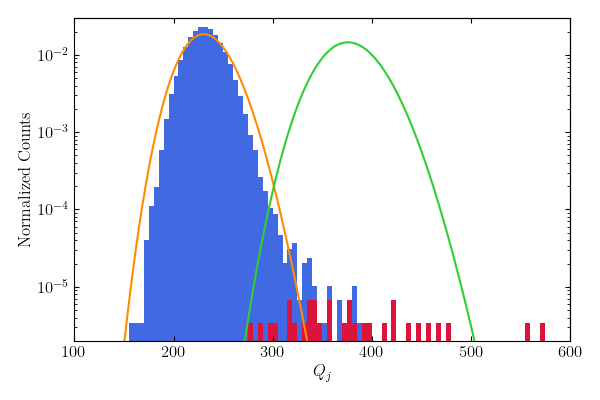}
\caption{\label{fig:dist}%
    \emph{Left}: Distribution of $Q_0$ [see \eqref{Q0}] in the DPDM analysis for frequency range $0.1205\Hz\leq f\leq0.1279\Hz$ (which corresponds to a single approximate coherence time $T_\mathrm{coh}$).  In blue, we show a histogram of $Q_0$ values in this frequency range.  In orange, we fit this distribution to a $\chi^2$-distribution, finding a best-fit number of degrees of freedom $\nu_\mathrm{fit}=362.8$.  In green, we show a $\chi^2$-distribution with the naive $6K_0=378$ degrees of freedom.  All frequencies with sufficiently large $Q_0$ (shown in red) are considered DM signal candidates.  \emph{Right}: Distribution of $Q_j$ [see \eqref{Qj}] for our $j=1$ subset and the same frequency range as the left plot.  We again attempt to fit the distribution to a $\chi^2$-distribution (orange) and show a $\chi^2$-distribution with $6K_0$ degrees of freedom (green).  In red, we highlight the $Q_j$ values of the candidates.  Those with large values of $Q_j$ are ruled out as DM signals.  Note that the vertical axes of normalized counts are logarithmic in both plots.}
\end{figure*}

Before presenting our exclusion bounds, we first identify and evaluate any potential DM signal candidates.  The key quantities for assessing the significance of a DM candidate are the $z_{(i)k}$ variables.  \eqref[s]{DPDM_likelihood_Z} and (\ref{eq:axion_likelihood_Z}) indicate that in the absence of a signal ($\varepsilon=0$ or $g_{a\gamma}=0$, respectively), the real and imaginary parts of each component of $z_{(i)k}$ should be Gaussian variables with mean zero and variance 1/2.  We can therefore combine them into a single statistic
\begin{equation}\label{eq:Q0}
    Q_0=2\sum_{(i,)k}|z_{(i)k}|^2,
\end{equation}
which should follow a $\chi^2$-distribution with $6K_0$ degrees of freedom in the DPDM case and $2K_0$ degrees of freedom in the axion case, where $K_0$ is the number of coherence chunks in the total dataset (which depends on the approximate coherence time $T_\mathrm{coh}$).  In actuality, we find that $Q_0$ follows a $\chi^2$-distribution with slightly fewer degrees of freedom.  For each $T_\mathrm{coh}$, we compute $Q_0$ for all frequencies in the corresponding frequency range and fit this computed distribution of $Q_0$ to a $\chi^2$-distribution to determine the appropriate number of degrees of freedom $\nu_\mathrm{fit}$.  In the left plot of \figref{dist}, we show such a distribution of $Q_0$ for a given coherence time $T_\mathrm{coh}$ in the DPDM analysis and compare it to a fitted $\chi^2$-distribution, as well as a $\chi^2$-distribution with $6K_0$ degrees of freedom.

Once the distribution of $Q_0$ has been fit, they can then be converted into $p$-values via%
\footnote{\label{ftnt:conservative}%
It is clear from the left plot of \figref{dist} that the high-$Q_0$ end of the distribution is not well described by a $\chi^2$-distribution with $\nu_\mathrm{fit}$ degrees of freedom.  The estimate of $p_0$ in \eqref{p0} is therefore not accurate when $p_0$ is small (which is the regime of interest).  We note however that it is a \emph{conservative} estimate in the sense that \eqref{p0} will always be smaller than the true $p_0$, and so any true candidate will still be identified.}
\begin{equation}\label{eq:p0}
    p_0=1-F_{\chi^2(\nu_\mathrm{fit})}[Q_0],
\end{equation}
where $F_{\chi^2(\nu)}$ is the cumulative distribution function of the $\chi^2$-distribution with $\nu$ degrees of freedom.  We consider there to be a naive DM signal candidate (with 95\% global confidence) at a given frequency if $p_0$ is below the threshold $p_\mathrm{crit}$ defined by
\begin{equation}
    (1-p_\mathrm{crit})^{N_f}=0.95,
\end{equation}
where $N_f$ is the number of frequencies in our range of consideration $10^{-3}\Hz\leq f\leq 0.98\Hz$.  In this work, we have $N_f\approx6.7\times10^6$, resulting in $p_\mathrm{crit}\approx7.7\times10^{-9}$.  Our candidate identification procedure identified 2931 DPDM signal candidates and 1186 axion signal candidates.

We now wish to evaluate the robustness of these naive signal candidates.  In \citeR[s]{Fedderke:2021iqw,Arza_2022}, the identified signal candidates were evaluated by reperforming the analysis on subsets of the data (either a shorter temporal duration or a subset of the stations) and considering whether the signal persisted.  Importantly, this was accomplished by comparing the inferred values of the coupling $\varepsilon$ or $g_{a\gamma}$ between the full and subset analyses.  The validation procedure used in \citeR[s]{Fedderke:2021iqw,Arza_2022} therefore did not utilize the full coherence-chunk-to-coherence-chunk phase/directional information contained in the $z_{(i)k}$ variables, but rather reduced all the information to a single variable.  The analysis in this work identified many more signal candidates than the $\mathcal O(30)$ identified in \citeR[s]{Fedderke:2021iqw,Arza_2022}, and so we require a significantly more powerful validation procedure.  In this work, we therefore design a new validation procedure that utilizes the full information contained in the $z_{(i)k}$ variables.

In the validation procedure for this work, we reperform the analysis on four random subsets of stations.  The relevant quantities that the DPDM subset analyses yield are $\bm Z_{k,j}$, $S_{k,j}$ and $V_{k,j}$ (where $j=1,2,3,4$ indexes the station subset used), while the relevant quantities from the axion analyses are $z_{k,j}$ and $s_{k,j}$.  In the DPDM case, we then combine these quantities with the corresponding quantities from the full-dataset analysis to form the variables
\begin{equation}\label{eq:DPDM_zeta}
    \bm\zeta_{k,j}=V_kS_k^{-1}\bm Z_k-V_{k,j}S_{k,j}^{-1}\bm Z_{k,j}.
\end{equation}
The key property of these variables is that any contributions to $\bm Z_k$ and $\bm Z_{k,j}$ from a true DPDM signal should cancel.  From \eqref[s]{dk} and (\ref{eq:DPDM_likelihood_Z}), we see that a DPDM signal should manifest as
\begin{align}
    \bm Z_k&\rightarrow\bm Z_k+\varepsilon S_kV_k^\dagger\bm c_k^*\\
    \bm Z_{k,j}&\rightarrow\bm Z_{k,j}+\varepsilon S_{k,j}V_{k,j}^\dagger\bm c_k^*.
\end{align}
Importantly, the $\varepsilon$ and $\bm c_k$ parameters appearing in both these expressions are the same because they are properties of the DPDM itself, not the stations used in the analysis.  Therefore the variables $\bm\zeta_{k,j}$ should exhibit the same mean-zero Gaussian distribution whether or not a true DM signal is present.  Anomalously large values of $\bm\zeta_{k,j}$ are instead indicative of a false DM signal candidate, which appears differently in the subsets as compared to the full dataset.  The analogous variables we define in the axion case are
\begin{equation}
    \zeta_{k,j}=\frac{z_k}{s_k}-\frac{z_{k,j}}{s_{k,j}},
\end{equation}
which should again cancel the contributions of any true axion DM signal.

Since we expect $\bm\zeta_{k,j}$ (or $\zeta_{k,j}$ in the axion case) to be mean-zero Gaussian variables, we can attempt to combine them into a $\chi^2$-statistic as we did in the candidate identification.  Based on \eqref[s]{DPDM_likelihood_Z} and (\ref{eq:DPDM_zeta}), the covariance matrix of $\bm\zeta_{k,j}$ should be
\begin{align}
    \Xi_{k,j}&\equiv2\langle\bm\zeta_{k,j}\bm\zeta_{k,j}^\dagger\rangle\\
    &=V_kS_k^{-2}V_k^\dagger+V_{k,j}S_{k,j}^{-2}V_{k,j}^\dagger.
\end{align}
We may then use $\Xi_{k,j}$ to normalize $\bm\zeta_{k,j}$ and form a $\chi^2$-squared statistic
\begin{equation}\label{eq:Qj}
    Q_j=2\sum_k\left|B_{k,j}^{-1}\bm\zeta_{k,j}\right|^2,
\end{equation}
where $\Xi_{k,j}=B_{k,j}B_{k,j}^\dagger$.  In the axion case, the analogous $\chi^2$-statistic is given by
\begin{align}
    \Xi_{k,j}&=s_k^{-2}+s_{k,j}^{-2},\\
    Q_j&=2\sum_k\left|\frac{\zeta_{k,j}}{\sqrt{\Xi_{k,j}}}\right|^2.
\end{align}

The right plot of \figref{dist} shows the distribution of $Q_j$ for our $j=1$ subset (and the same frequency range as the left plot).  Unlike the case of $Q_0$, the distribution of $Q_j$ does not match a $\chi^2$-distribution (either with $6K_0$ degrees of freedom or with a best-fit degrees of freedom $\nu_\mathrm{fit}$), so we cannot convert $Q_j$ into a $p$-value using a $\chi^2$-distribution, as we did in \eqref{p0}.  Instead, we must use the empirical distribution of $Q_j$ values.  That is, for each coherence time $T_\mathrm{coh}$, we compute $Q_j$ for all frequencies in the corresponding frequency range.  Then for each candidate frequency $f_p$, we compute its $p$-value as
\begin{equation}\label{eq:pj}
    p_j=1-\frac{n_{Q_j}[f_p]-2}{N_f[T_\mathrm{coh}]-1},
\end{equation}
where $N_f[T_\mathrm{coh}]$ denotes the number of frequencies in the range corresponding to $T_\mathrm{coh}$, and $n_{Q_j}[f_p]$ denotes the index of $f_p$ when the frequencies in this range are sorted in ascending order by their $Q_j$ values (so that $n_{Q_j}=1$ for the lowest value and $n_{Q_j}=N_f[T_\mathrm{coh}]$ for the highest).  The $-1$ in the denominator is present so that $f_p$ itself is not included in the distribution of $Q_j$ used to estimate the $p$-value of $f_p$.  \eqref{pj} is defined to be a conservative estimate of $p_j$; that is, it gives an upper bound on the true value of $p_j$.  For instance, \eqref{pj} will give $p_j=1/(N_f[T_\mathrm{coh}]-1)$ for the frequency with the largest $Q_j$ value (though its true value of $p_j$ may be smaller).%
\footnote{For most coherence times $T_\mathrm{coh}$, typically $N_f[T_\mathrm{coh}]\approx5.9\times10^3$, so that \eqref{pj} always gives an estimate $p_j\geq1.7\times10^{-5}$.  This technique would therefore not have been viable for estimating $p_0$, as we need to be able to estimate $p$-values smaller than $p_\mathrm{crit}$.  Instead, we must rely on a $\chi^2$-distribution, as in \eqref{p0}.  [See also footnote~\ref{ftnt:conservative}.]}

Using this procedure, each candidate can be re-evaluated against all four station subsets.  The resulting $p_j$ can be combined into a single $p$-value via Fisher's method~\cite{Fisher:1958iqe}
\begin{align}
    Q_\mathrm{full}&=-2\sum_{j=1}^4\ln p_j,\\
    p_\mathrm{full}&=1-F_{\chi^2(8)}(Q_\mathrm{full}).
\end{align}
These $p_\mathrm{full}$ summarize whether the DM signal identified in the full dataset is consistent with the subset analyses.  In \tabref{resampling}, we list all DPDM and axion DM candidates with $p_\mathrm{full}>10^{-3}$.  In addition to their $p_0$, $p_j$, and $p_\mathrm{full}$ values, we also translate their $p_0$ into a one-sided global significance
\begin{equation}
    \sigma(p_0)\equiv\sqrt2\,\mathrm{erfc}^{-1}\left[ 2\left(1-\left(1-p_0\right)^{N_{f}}\right)\right].
\end{equation}
Note that among the 2931 DPDM candidates and 1186 axion DM candidates we identified, only a single DPDM candidate ($f_{A'}=0.9511\Hz$) and a single axion DM candidate ($f_a=0.0303\Hz$) have $p_\mathrm{full}>0.01$.  Both of these candidates have weak global significance ($\sigma(p_0)<2$) and still exhibit tension with the subset analyses ($p_\mathrm{full}<0.05$).  We therefore deem that our search finds no robust DM candidates.  It may, however, be useful to check these weak candidates in future SNIPE Hunt data.  (The axion candidate lies below the frequency range considered in \citeR{Sulai_2023}, but may be probed with additional low-frequency data.)  Moreover, we also note that none of the signal candidates found in \citeR[s]{Fedderke:2021iqw,Arza_2022} appear in \tabref{resampling}, and so we reaffirm the claims of those works that none of their candidates were robust.

\begin{table}[t]
    \centering
    \begin{ruledtabular}
    \begin{tabular}{c|cc|cccc|c}
    $f_\DM$\,[Hz] & $p_0$ & $\sigma(p_0)$ & $p_1$ & $p_2$ & $p_3$ & $p_4$ & $p_\mathrm{full}$ \\\hline
    $0.1100$ & $7.5\times10^{-9}$   &   $1.7$   &   $0.070$   &   $0.205$   &   $0.010$   & $0.030$ & $0.002$ \\
    $0.3053$ & $3.5\times10^{-9}$ & $2.0$ & $0.085$ & $0.004$ & $0.388$ & $0.133$ & $0.005$ \\
    $0.6217$ & $2.8\times10^{-9}$ & $2.1$ & $0.069$ & $0.010$ & $0.096$ & $0.132$ & $0.003$ \\
    $0.9511$ & $5.6\times10^{-9}$ & $1.8$ & $0.073$ & $0.451$ & $0.422$ & $0.024$ & $0.042$ \\
    $0.9703$ & $2.5\times10^{-9}$ & $2.1$ & $0.064$ & $0.120$ & $0.021$ & $0.068$ & $0.004$ \\
    \hline
    $0.0303$ & $5.2\times10^{-9}$ & $1.8$ & $0.194$ & $0.079$ & $0.039$ & $0.607$ & $0.045$ \\
    $0.0351$ & $2.3\times10^{-9}$ & $2.2$ & $0.147$ & $0.001$ & $0.229$ & $0.183$ & $0.002$ \\
    $0.2630$ & $1.1\times10^{-9}$ & $2.4$ & $0.404$ & $0.001$ & $0.104$ & $0.350$ & $0.003$
    \end{tabular}
    \end{ruledtabular}
    \caption{\label{tab:resampling}%
    List of DPDM (top section) and axion DM (bottom section) candidates with $p_\mathrm{full}>10^{-3}$.  For each candidate we show: its frequency, $f_\DM$; its local $p$-value in the full dataset, $p_0$; its global significance in the full dataset, $\sigma(p_0)$; its $p$-value for agreement with each individual subset, $p_j$; and the combined $p$-value for all subset analyses, $p_\mathrm{full}$.  Note that no candidates exhibit good agreement with the subset analyses ($p_\mathrm{full}>0.05$), and even the single DPDM candidate and single axion DM candidate with $p_\mathrm{full}>0.01$ have weak global significance ($\sigma(p_0)<2$).
    }
\end{table}

Finally, we note that we have validated our analysis pipeline and candidate identification/evaluation procedure by injecting a mock signal into our dataset.  In the DPDM case, we injected an exactly monochromatic signal of the form in \eqref{DPDM_signal} directly into the SuperMAG data, while in the axion DM case, we inject an exactly monochromatic signal of the form in \eqref{axion_signal}.%
\footnote{By monochromatic, we mean that the injected signal follows \eqref{DPDM_signal} or (\ref{eq:axion_signal}) for the entire duration of the dataset, in contrast to the finite coherence-time behavior described in \secref{signal}.  \citeR{Fedderke:2021iqw} validated its analysis pipeline using a signal with finite coherence time.  (\citeR{Fedderke:2021iqw} injected this signal into the time series $X^{(n)}$ rather than directly into the data, as was done in this work.)  As the analysis in this work is similar, especially in its handling of the finite coherence time, here we simply validate by injecting a monochromatic signal.}
In both cases, our candidate identification procedure identifies the exact same set of signal candidates as the original analysis, with the addition of the frequency $f_*$ at which the DM signal was injected, as well as its reflection across the Nyquist frequency $1\Hz-f_*$.  These additional candidates all exhibit large $p_\mathrm{full}\gtrsim30\%$, indicating that these candidates manifest similarly in the subset analyses as in the full dataset.

\begin{figure*}[t]
\includegraphics[width=0.49\textwidth]{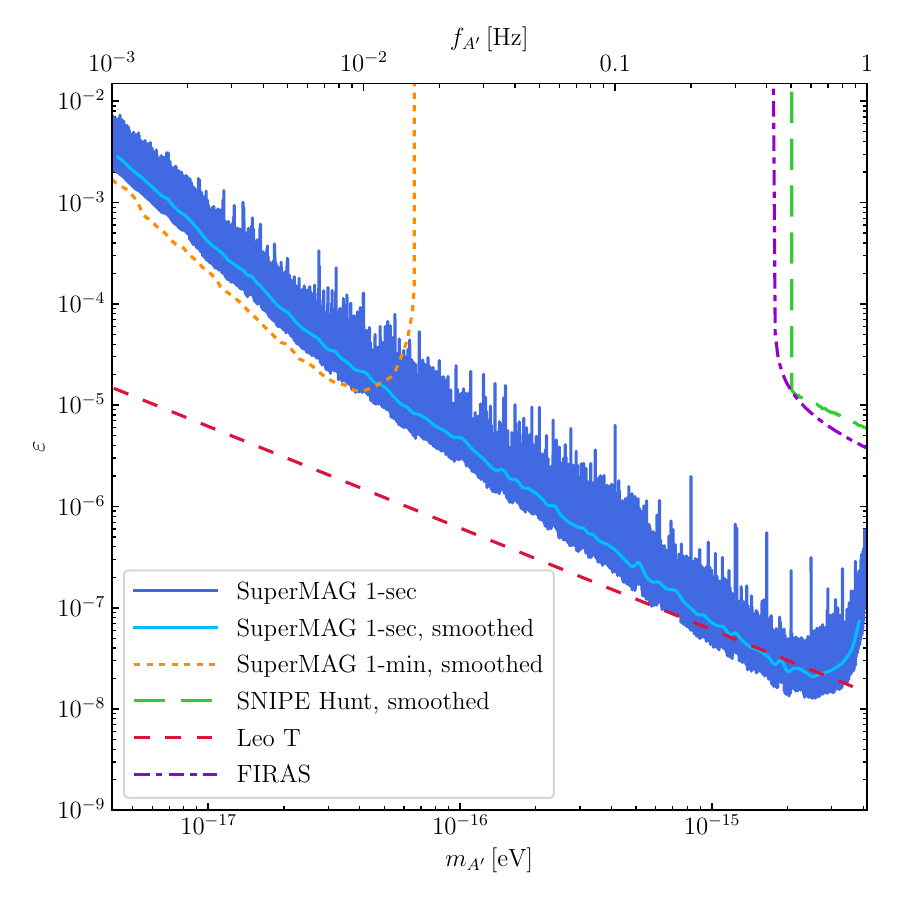}
\includegraphics[width=0.49\textwidth]{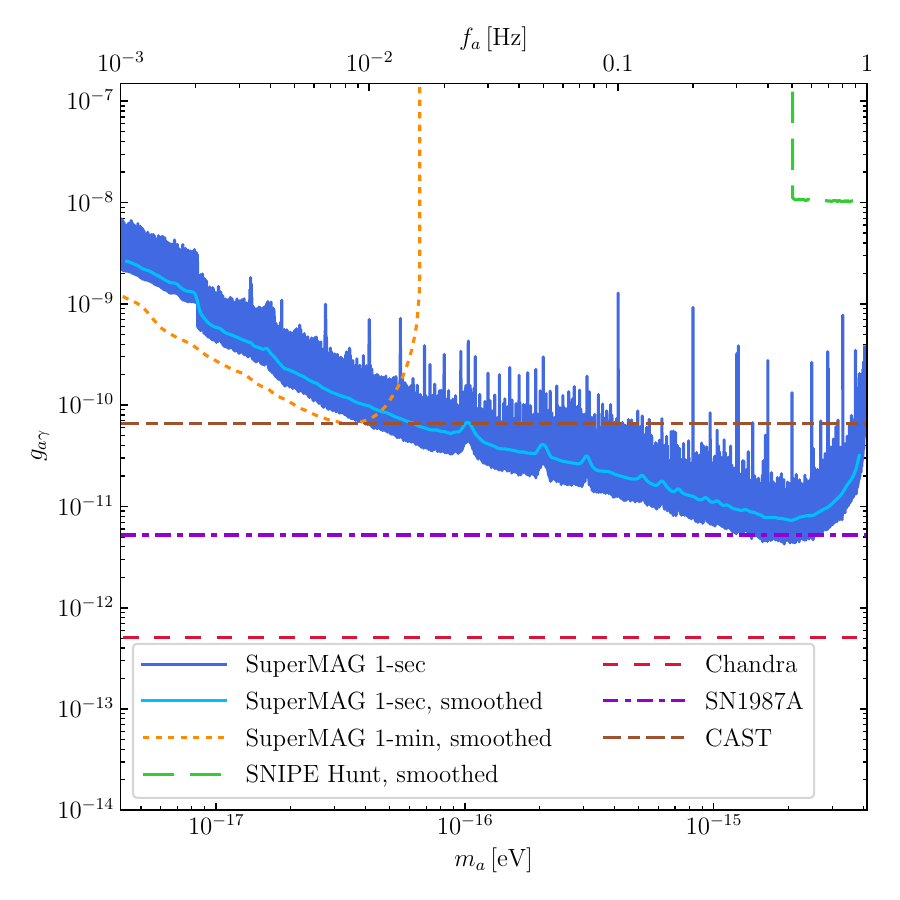}
\caption{\label{fig:bounds}%
    Constraints on DPDM (left) and axion DM (right) based on our search of the SuperMAG high-fidelity dataset.  In solid darker blue, we show our 95\%-credible upper limits on $\varepsilon$ and $g_{a\gamma}$, respectively, as a function of DM mass (corresponding Compton frequency is shown on the upper axis).  These constraints appear as bands due to frequency-to-frequency fluctuations, and also exhibit a number of narrow spikes, which correspond to potential DM signals.  We evaluate these candidates in \secref{candidates} and find that none constitute robust evidence of DM.  In solid lighter blue, we also show sliding averages of the limits, which are computed by smoothing with a Gaussian filter on a log scale [see \eqref{limit_smoothing}].  In various dashed colors, we show existing constraints on DPDM and axion DM (see \secref{conclusion} for descriptions).  Our constraints are the leading direct probes of both these DM candidates in the mass range $7\times10^{-17}\eV\lesssim m_\DM\lesssim4\times10^{-15}\eV$, and our DPDM constraint even surpasses the leading astrophysical constraint in a narrow mass range around $m_{A'}\approx2\times10^{-15}\eV$.
    }
\end{figure*}

\section{Results and conclusion}
\label{sec:conclusion}

Having found no robust evidence for a DM signal, we move on to present and discuss the constraints on DPDM and axion DM derived from our analysis.  In \figref{bounds}, we show our resulting constraints on $\varepsilon$ and $g_{a\gamma}$ in darker blue (with the degradation factors in \eqref[s]{DPDM_degrade} and (\ref{eq:axion_degrade}) applied).  This bound appears as a band due to frequency-bin-to-frequency-bin fluctutations, and also exhibits several narrow peaks, corresponding to the DM signal candidates identified via the procedure in \secref{candidates}.  In lighter blue, we show a sliding average of this bound.  We compute this average using a Gaussian filter, similar to the one used in \eqref{PSDsmoothing} in the noise estimation procedure, but here we perform the smoothing in log space
\begin{align}\label{eq:limit_smoothing}
    \tilde\varepsilon(f_p)&=\sum_q\mathcal K\left(\frac{\log f_{q+1}+\log f_q}2-\log f_p;\sigma_{\log}\right),\nl
    \qquad\cdot\Delta\log f_q\cdot\frac{\hat\varepsilon(f_{q+1})+\hat\varepsilon(f_q)}2\\
    \Delta\log f_q&\equiv\log f_{q+1}-\log f_q,
\label{eq:Deltalog}\end{align}
where $f_p$ runs over all the frequencies at which we perform our analysis and $\mathcal K$ is a Gaussian filter [as in \eqref{filter}] with window size $\sigma_{\log}=0.02$.  (A similar expression is used to compute a smoothed limit $\tilde g_{a\gamma}$ from $\hat g_{a\gamma}$.)

\figref{bounds} also shows a number of existing constraints on DPDM and axion DM for comparison.  The DPDM constraints include limits derived from: the SuperMAG low-fidelity dataset~\cite{Fedderke:2021iqw}; magnetometer measurements made by the SNIPE Hunt collaboration~\cite{Sulai_2023} (we plot the smoothed versions of both of these limits); heating of the dwarf galaxy Leo T~\cite{Wadekar_2021}; and non-observation of CMB-photon conversion into (non-DM) dark photons by the FIRAS instrument~\cite{Caputo_2020}.  The limit derived in this work represents the strongest direct constraint on DPDM for masses $7\times10^{-17}\eV\lesssim m_{A'}\lesssim4\times10^{-15}\eV$.  Moreover, it is even competitive with the leading astrophysical constraint from Leo T near $m_{A'}\approx2\times10^{-15}\eV$.

The axion constraints shown in the right plot of \figref{bounds} include limits derived from: the SuperMAG low-fidelity dataset~\cite{Arza_2022}; SNIPE Hunt; X-ray observations of the quasar H1821+643 from the Chandra telescope~\cite{Sisk_Reynes_2021}; non-observation of gamma rays in coincidence with SN1987A~\cite{Hoof_2023}; and the CAST helioscope search for solar axions~\cite{Anastassopoulos:2017ftl}.  The limit derived in this work is again the strongest direct constraint on axion DM for masses $7\times10^{-17}\eV\lesssim m_a\lesssim4\times10^{-15}\eV$, surpassing even the constraints from CAST in this mass range.  Moreover, it is competitive with the SN1987A constraint near $m_a\approx2\times10^{-15}\eV$.

This work employed the method proposed in \citeR[s]{Fedderke:2021rys,Fedderke:2021iqw,Arza_2022} to utilize unshielded magnetometer data maintained by the SuperMAG collaboration in order to constrain DPDM and axion DM.  The results derived in this work cover a gap in mass range between the limits from the SuperMAG low-fidelity dataset and the limits from the SNIPE Hunt collaboration.  While our search did not identify any robust DM candidates, it demonstrates the strength of terrestrial magnetic-field measurements as leading direct probes of ultralight DM parameter space.  This motivates further measurements by collaborations such as SNIPE Hunt, or further efforts to interpret existing data in order to detect DPDM or axion DM.

\acknowledgments

We thank Ariel Arza, Michael Fedderke, Peter Graham, and Derek Jackson Kimball for their helpful discussions.  S.K. is supported in part by the U.S. Department of Energy, Office of Science, National Quantum Information Science Research Centers, Superconducting Quantum Materials and Systems Center (SQMS) under contract number DE-AC02-07CH11359.  The code used for this research is made publicly available through Github~\cite{github} under CC-BY-NC-SA.

Some of the computing for this project was performed on the Sherlock cluster. We would like to thank Stanford University and the Stanford Research Computing Center for providing computational resources and support that contributed to these research results.

We gratefully acknowledge the SuperMAG Collaboration for maintaining and providing the database of ground magnetometer data that were analyzed in this work.
SuperMAG receives funding from NSF Grant Nos.~ATM-0646323 and AGS-1003580, and NASA Grant No.~NNX08AM32G S03.

We acknowledge those who contributed data to the SuperMAG Collaboration: 
INTERMAGNET, Alan Thomson; 
CARISMA, PI Ian Mann; 
CANMOS, Geomagnetism Unit of the Geological Survey of Canada; 
The S-RAMP Database, PI K.~Yumoto and Dr.~K.~Shiokawa; 
The SPIDR database; AARI, PI Oleg Troshichev; 
The MACCS program, PI M.~Engebretson; 
GIMA; 
MEASURE, UCLA IGPP and Florida Institute of Technology; 
SAMBA, PI Eftyhia Zesta; 
210 Chain, PI K.~Yumoto; 
SAMNET, PI Farideh Honary; 
IMAGE, PI Liisa Juusola; 
Finnish Meteorological Institute, PI Liisa Juusola; 
Sodankylä Geophysical Observatory, PI Tero Raita; 
UiT the Arctic University of Norway, Troms\o\ Geophysical Observatory, PI Magnar G.~Johnsen; 
GFZ German Research Centre For Geosciences, PI J\"urgen Matzka; 
Institute of Geophysics, Polish Academy of Sciences, PI Anne Neska and Jan Reda; 
Polar Geophysical Institute, PI Alexander Yahnin and Yarolav Sakharov; 
Geological Survey of Sweden, PI Gerhard Schwarz; 
Swedish Institute of Space Physics, PI Masatoshi Yamauchi; 
AUTUMN, PI Martin Connors; 
DTU Space, Thom Edwards and PI Anna Willer; 
South Pole and McMurdo Magnetometer, PIs Louis J.~Lanzarotti and Alan T.~Weatherwax; 
ICESTAR; 
RAPIDMAG; 
British Antarctic Survey; 
McMac, PI Dr.~Peter Chi; 
BGS, PI Dr.~Susan Macmillan; 
Pushkov Institute of Terrestrial Magnetism, Ionosphere and Radio Wave Propagation (IZMIRAN); 
MFGI, PI B.~Heilig; 
Institute of Geophysics, Polish Academy of Sciences, PI Anne Neska and Jan Reda; 
University of L’Aquila, PI M.~Vellante; 
BCMT, V.~Lesur and A.~Chambodut; 
Data obtained in cooperation with Geoscience Australia, PI Marina Costelloe; 
AALPIP, co-PIs Bob Clauer and Michael Hartinger; 
SuperMAG, PI Jesper W.~Gjerloev; 
data obtained in cooperation with the Australian Bureau of Meteorology, PI Richard Marshall.

We thank INTERMAGNET for promoting high standards of magnetic observatory practice~\cite{INTERMAGNETwebsite}.

\appendix

\section{Vector spherical harmonics}
\label{app:VSH}

This appendix, which defines the VSH conventions used in this work, is reproduced from \citeR{Fedderke:2021rys} with minor modifications for the convenience of the reader.

The vector spherical harmonics are defined in terms of the scalar spherical harmonic $Y_{\ell m}$ by the relations
\begin{align}
\bm{Y}_{\ell m} &= Y_{\ell m}\rhat, &
\bm{\Psi}_{\ell m} &= r\bm{\nabla} Y_{\ell m}, &
\bm{\Phi}_{\ell m} &= \bm{r}\times\bm{\nabla} Y_{\ell m},
\end{align}
where $\rhat$ is the unit vector in the direction of $\bm{r}$.  Thus $\bm{Y}_{\ell m}$ points radially, while $\bm{\Psi}_{\ell m}$ and $\bm{\Phi}_{\ell m}$ point tangentially.  Some of their relevant properties (and our phase conventions) are
\begin{align}
\bm{Y}_{\ell,-m}&=(-1)^m\bm{Y}_{\ell m}^*,\\
\bm{\Psi}_{\ell,-m}&=(-1)^m\bm{\Psi}_{\ell m}^*,\\
\bm{\Phi}_{\ell,-m}&=(-1)^m\bm{\Phi}_{\ell m}^*, \label{eq:phaseConventionPointer}\\
\bm{Y}_{\ell m}\cdot\bm{\Psi}_{\ell m}&=\bm{Y}_{\ell m}\cdot\bm{\Phi}_{\ell m}=\bm{\Psi}_{\ell m}\cdot\bm{\Phi}_{\ell m}=0,
\end{align}
\begin{align}
\int d\Omega\,\bm{Y}_{\ell m}\cdot \bm{Y}_{\ell'm'}^*&=\delta_{\ell\ell'}\delta_{mm'},\\
\int d\Omega\,\bm{\Psi}_{\ell m}\cdot\bm{\Psi}_{\ell'm'}^*&=\int d\Omega~\Phi_{\ell m}\cdot\Phi_{\ell'm'}^*\nonumber \\
	&=\ell(\ell+1)\delta_{\ell\ell'}\delta_{mm'},\\
\int d\Omega\,\bm{Y}_{\ell m}\cdot\bm{\Psi}_{\ell'm'}^*&=\int d\Omega\,\bm{Y}_{\ell m}\cdot\bm{\Phi}_{\ell'm'}^*\nonumber \\
	&=\int d\Omega\,\bm{\Psi}_{\ell m}\cdot\bm{\Phi}_{\ell'm'}^*=0.\label{eq:orthogonality}
\end{align}
The explicit expressions for the spherical harmonics that are relevant to the DPDM signal [see \eqref{DPDM_signal}] are
\begin{align}
\bm{\Phi}_{1,-1}(\bm{r}) &=\sqrt{\frac3{8\pi}}e^{-i\phi}(i\thetahat+\cos\theta\phihat), \label{eq:Phi1m1}\\
\bm{\Phi}_{10}(\bm{r})&=-\sqrt{\frac3{4\pi}}\sin\theta\phihat,\label{eq:Phi10}\\
\bm{\Phi}_{11}(\bm{r})&=\sqrt{\frac3{8\pi}}e^{i\phi}(i\thetahat-\cos\theta\phihat)\label{eq:Phi1p1},
\end{align}
where $\thetahat$ and $\phihat$ are unit vectors in the directions of increasing $\theta$ and $\phi$.

Recall that, as written here, the spherical coordinate $\phi$ coincides with the definition of longitude; however, the spherical coordinate $\theta$ is not the latitude: $\theta$ increases from $\theta = 0$ at the Geographic North Pole (latitude $+90^\circ$), to $\theta = \pi/2$ on the Equator (latitude $0^\circ$), to $\theta = \pi$ at the Geographic South Pole (latitude $-90^\circ$).

\section{Parameter validation}
\label{app:noise}

In this appendix, we validate our choice of parameters by examining how they affect the statistics of the noise.  In particular, we will first evaluate the choice of stationarity timescale by considering its effect on the noise spectra $S^a_{mn}(f)$.  Then we will consider the choice of Gaussian filter window $\sigma$ used in \eqref{filter} by examining the Gaussianity of the final analysis variables $z_{(i)k}$.

\subsection{Stationarity timescale}
\label{app:stationarity}

The stationarity timescale is the timescale within which the noise in the time series $X^{(n)}$ is assumed to be constant.  In \citeR[s]{Fedderke:2021iqw,Arza_2022}, this timescale was chosen to be a calendar year, as it was observed that many stations become active/inactive at the beginning of calendar years, and so the noise in $X^{(n)}$ varies appreciably between one year and the next.  As the analysis in this work deals with higher frequencies, it is natural to think that a shorter stationarity timescale may be appropriate.  Again it will be prudent to choose a stationarity timescale so that stations are likely to become active/inactive on the boundary between stationarity periods.  Natural candidates may be: calendar year, calendar month, week, or day.  We will investigate the merits of these possible choices in this subsection.

The stationarity timescale is relevant to two parts of our analysis.  It is both the timescale for which we choose the weights $w^{(n)}_i$ via \eqref[s]{weights_theta} and (\ref{eq:weights_phi}) and also the timescale for which the noise spectra $S^a_{mn}(f)$ are calculated via \eqref{PSDsmoothing}.  Both of these contexts will favor the shorter weekly or daily timescales.  For computational reasons, it is simpler to compute noise spectra for fewer stationarity periods, and so we will choose the longer timescale of a week.

Let us first consider the latter context.  The purpose of calculating the noise spectra $S^a_{mn}(f)$ is to ultimately determine the covariance matrix $\Sigma_k$ for a given coherence chunk.  This covariance matrix is computed by summing over all stationarity periods that overlap with the coherence chunk, as in \eqref[s]{DPDM_sigma} and (\ref{eq:axion_sigma}).  So long as the coherence chunk of interest spans a duration much longer than a year, the difference between summing over all overlapping years vs. all overlapping months/weeks/days is negligible.  However, if the coherence chunk is much shorter than a year, then the former will sum over a significantly longer timespan.  In particular, if transient events occur in the same calendar year as (but outside of) the coherence chunk, these will affect the noise calculation in the former case but not in the latter case.  Since the coherence time in this work can be as short as $T_\mathrm{coh}\sim10^6/(1\Hz)\sim12\,\mathrm{days}$, this makes a year (and likewise any timescale longer than 12 days) a bad candidate for the stationarity timescale.  A week or a day are instead more acceptable timescales in this context.

The stationarity timescale plays a much more significant role in the selection of weights.  The purpose of the weighting in \eqref{Xn} is to downweight noisier stations.  This is done on the stationarity timescale, so if a transient event of short duration occurs, then the station will be downweighted for the entire stationarity period in which that transient occurs.  Moreover, the extent of the downweighting is determined by the white noise level of the station within the stationarity period in question [see \eqref[s]{weights_theta} and (\ref{eq:weights_phi})].  This means that transient events will be downweighted less when the stationarity timescale is larger because the contribution of the transient to the overall white noise level of the station will be less.  For both of these reasons, it is preferable to have a shorter stationarity timescale.

\begin{figure*}[t]
\includegraphics[width=0.99\textwidth]{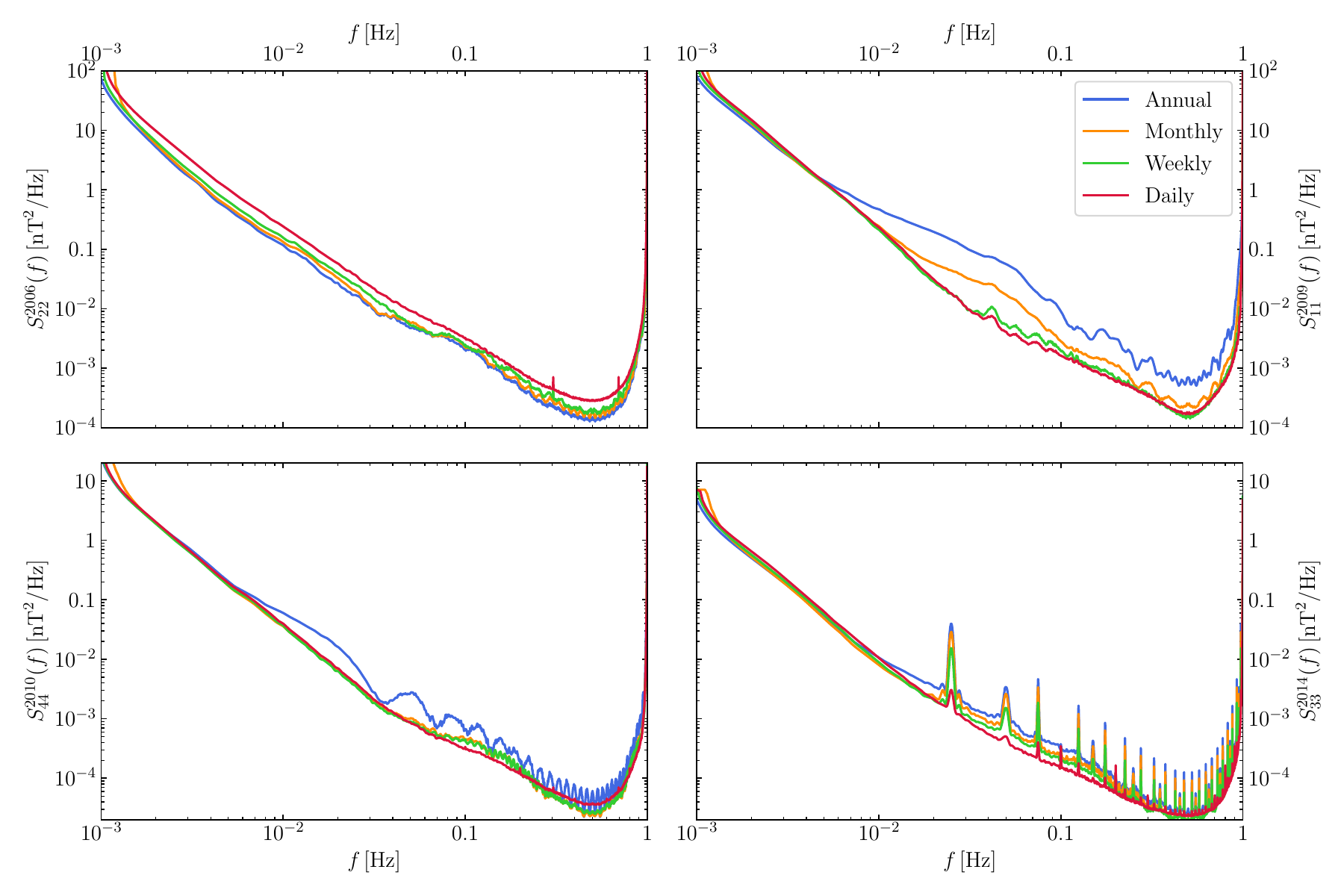}
\caption{\label{fig:stationarity}%
    Dependence of noise spectra $S^{a'}_{nn}$ (for various years $a'$ and time series $X^{(n)}$) in the DPDM analysis on stationarity timescale.  We show the spectra as calculated using stationarity timescale (and corresponding window size $\sigma$): calendar year ($\sigma=10^4$, blue), calendar month ($\sigma=1000$, orange), week ($\sigma=200$, green), and day ($\sigma=30$, red).  [See \eqref{annual_spectra} for how the spectrum for a year is defined when the stationarity timescale is not a year.]  These large window sizes (corresponding to a frequency width of roughly $3\times10^{-4}\,\mathrm{Hz}$) are used for illustrative purposes, as they lead to spectra that are much smoother than those used in the actual analysis.  Note that the lowest noise levels are typically achieved with weekly or daily timescales.}
\end{figure*}

The effect of the stationarity timescale on the noise in the time series is demonstrated in \figref{stationarity}.  In this figure, we show the (diagonal) noise spectra $S^{a'}_{nn}$ for a given year $a'$ and time series $X^{(n)}$ [in the DPDM case], calculated with different stationarity timescales (for both the selection of weights and noise estimation).  In the ``annual" case, this is simply the noise spectrum calculated via \eqref{PSDsmoothing}.  In the other cases, the noise spectra are computed on the shorter stationarity timescale and then summed over several timescales to determine the spectrum for the full year, similar to the sum in \eqref{Xk_variance}.  Explicitly, to determine the noise spectrum for year $a'$, we calculate
\begin{equation}\label{eq:annual_spectra}
    S_{nn}^{a'}(f)=\frac1{T^{a'}}\sum_a T_{a'}^a \cdot S_{nn}^a(f),
\end{equation}
where $T^{a'}$ denotes the duration of year $a'$ and $T_{a'}^a$ denotes the duration of the overlap between stationarity period $a$ and year $a'$.  When computing the noise spectra for \figref{stationarity}, we utilize a large window size $\sigma$, corresponding in all cases to a frequency width of roughly $3\times10^{-4}\,\Hz$ (see figure caption for exact values of $\sigma$).  This leads to much smoother noise spectra with significantly lower variance than those that are used in the actual analysis.  As the purpose of \figref{stationarity} is only to demonstrate how the mean behavior varies with different stationarity timescales, we find this extra smoothing to be useful for illustration.

\figref{stationarity} demonstrates that the stationarity timescale has little effect on low-frequency noise, but rather its effect is larger at higher frequencies, where noise levels are lower and spurious features in the data can have more influence.  In most cases, we see that the lowest noise levels are achieved using a stationarity timescale of a week or a day, with little difference between the two in the absence of spurious features.  The lower right plot demonstrates that shorter stationarity timescales reduce the effect of spurious features.  As explained above, this occurs because transient events are more efficiently downweighted when the stationarity timescale is shorter.  While this reasoning may make a daily timescale slightly more optimal, we instead opt for the more computationally tractable weekly timescale.

\subsection{Filter window}
\label{app:window}

The second analysis parameter that requires validation is the window size $\sigma$ used in the Gaussian filter defined in \eqref{filter}.  This window determines the scale over which the noise spectra are smoothed in the noise estimation procedure.  In our analysis, we are searching for signals of width $10^{-6}f_\DM$ in frequency space.  We identify such signals by comparing them to the local noise.  Our smoothing scale should therefore be much larger than the width of a potential signal so that we are not comparing the signal to a noise estimated from the signal itself.  Moreover, $\sigma$ should be large enough that our smoothing procedure averages over enough frequency bins to get an accurate estimate of the noise.  On the other hand, if we smooth over a window significantly larger than the width of the actual features in the noise spectrum, we will end up comparing potential signals to noise levels that do not accurately characterize the noise in their local frequency range.  This will manifest as improper statistics of the $z_{(i)k}$ variables, which according to our analysis should be Gaussian with mean zero and $\langle|z_{(i)k}|^2\rangle=1$.

\begin{figure}[!t]
\includegraphics[width=\columnwidth]{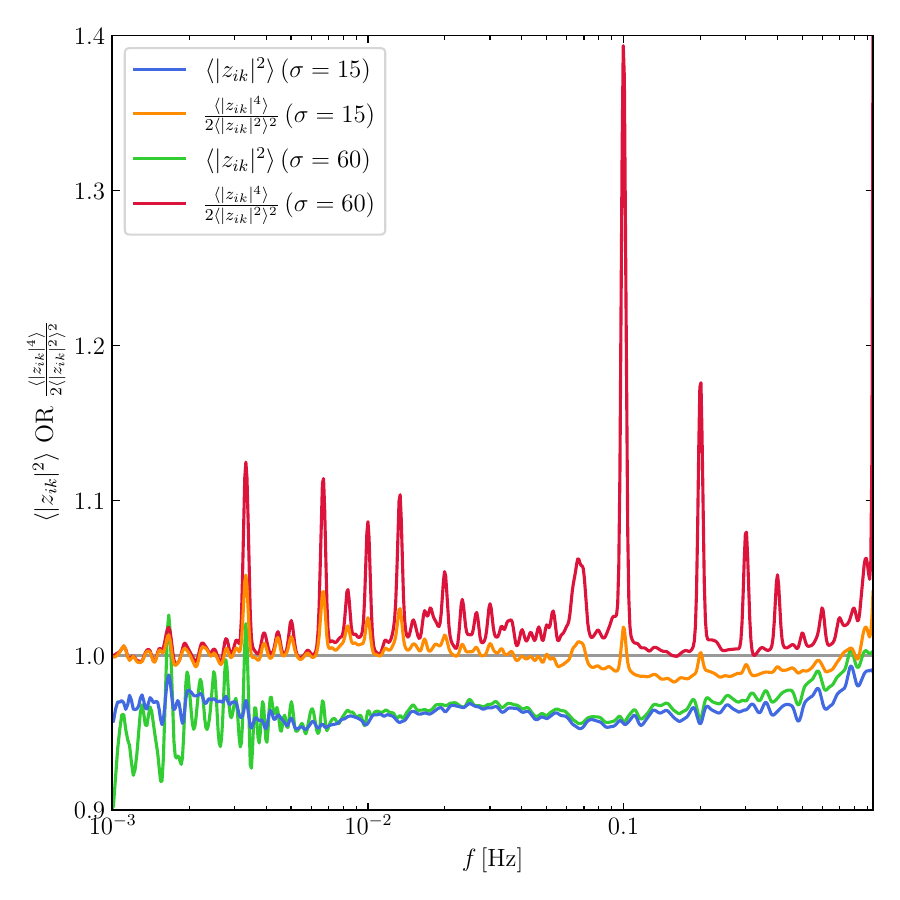}
\caption{\label{fig:gaussianity}%
    Dependence of the statistics of $z_{ik}$ in the DPDM analysis on the Gaussian filter window size $\sigma$ [see \eqref{filter}].  In blue ($\sigma=15$ frequency bins) and green ($\sigma=60$), we show $\langle|z_{ik}|^2\rangle$, while in orange ($\sigma=15$) and red ($\sigma=60$), we show the ratio $\langle|z_{ik}|^4\rangle/2\langle|z_{ik}|^2\rangle^2$, all as a function of frequency (see \eqref{zaverage} for definition of averaging procedure).  If the noise is properly modeled, both should be equal to 1 (indicated by the grey line).  The smaller window size $\sigma=15$ exhibits much better statistics.  See also Fig.~11 of \citeR{Fedderke:2021iqw} for the analogous plot in that work.}
\end{figure}

In \figref{gaussianity}, we show the dependence of the statistics of $z_{ik}$ (in the DPDM case) on the filter window size.  We show both the two- and four-point statistics of $z_{ik}$ in the cases $\sigma=15$ frequency bins and $\sigma=60$ frequency bins (corresponding to frequency widths of $2.5\times10^{-5}\,\Hz$ and $10^{-4}\,\Hz$, respectively).  The average $\langle\cdot\rangle$ in this figure is computed in a similar manner to the procedure used to produce the ``smoothed" bounds in \figref{bounds}, that is, we smooth with a Gaussian filter in log-space
\begin{align}\label{eq:zaverage}
    \langle|z_{ik}|^m\rangle(f_p)&=\sum_q\mathcal K\left(\frac{\log f_{q+1}+\log f_q}2-\log f_p;\sigma_{\log}\right)\nl
    \cdot\frac{\Delta\log f_q}{3K(f_q)}\cdot\sum_{i,k}\frac{|z_{ik}(f_{q+1})|^m+|z_{ik}(f_q)|^m}2,
\end{align}
for $m=2,4$, where $f_p$ runs over all the frequencies at which we perform our analysis; $\mathcal K$ is a Gaussian filter with window $\sigma_{\log}=0.02$; $\Delta\log f_q$ is as in \eqref{Deltalog}; $z_{ik}(f_p)$ are our analysis variables calculated for $f_{A'}=f_p$; and $K(f_p)$ is the number of coherence chunks for freqeuncy $f_p$ (which will differ for frequencies that have different approximate coherence times).  In \figref{gaussianity}, we show the quantity $\langle|z_{ik}|^2\rangle$ to demonstrate the two-point statistics and the ratio $\langle|z_{ik}|^4\rangle/2\langle|z_{ik}|^2\rangle^2$ to demonstrate the four-point statistics.  If the $z_{ik}$ variables are properly normalized and Gaussian, both of these quantities should be equal to 1.  Note that Fig.~11 of \citeR{Fedderke:2021iqw} shows similar statistics for the analysis of the low-fidelity dataset, but the averaging in that work was done with a top-hat instead of a Gaussian filter.

\figref{gaussianity} demonstrates that a larger window size leads to poorly behaved statistics for the analysis variables.  In particular, the four-point statistics in the $\sigma=60$ case differ from their expected value by as much as $40\%$.  Many of the large deviations in \figref{gaussianity} are only of width $\sigma_{\log}$, which is indicative of an anomaly that may only span a few frequency bins (but is broadened by the Gaussian filter).  As alluded to earlier, the smaller $\sigma=15$ window size models the noise near these anomalies more locally, thereby leading to four-point statistics that are closer to expectation.  The two-point statistics are similar in both cases at higher frequencies, but at lower frequencies the larger window size leads to larger deviations in the two-point statistics.  This is because the noise exhibits $1/f^\alpha$-behavior (see \figref{stationarity}), so the noise changes much more rapidly at low frequencies, and therefore a smaller window size is more appropriate.  The $\sigma=15$ case exhibits $\mathcal O(5\%)$ deviations in the two-point statistics and even smaller deviations in the four-point statistics, which indicates that our noise modeling is sufficiently accurate with this choice of window size.

\bibliographystyle{JHEP}
\bibliography{references.bib}

\end{document}